\documentclass[12pt]{article}
\usepackage[utf8]{inputenc}

\usepackage[left=1.5cm, right=1.5cm, top=1.785cm, bottom=2.0cm]{geometry}
\usepackage[dvipsnames,table,xcdraw]{xcolor}
\usepackage{mathtools}
\usepackage{float}
\usepackage{array}
\usepackage{multirow}

\usepackage{hyperref}
\usepackage[affil-it]{authblk}
\title{Thermophysical properties of $n$-hexadecane: Combined Molecular Dynamics and experimental investigations}
\date{}

\usepackage[numbers]{natbib}
\usepackage{graphicx}
\usepackage{caption}
\usepackage{subcaption}
\usepackage{soul}
\usepackage{url}
\usepackage{amsmath} 

\author[a,*]{L. Klochko}
\author[a]{J. Noel}
\author[a]{N. R. Sgreva}
\author[a]{S. Leclerc}
\author[a]{C. M{\'e}tivier}
\author[a]{D. Lacroix}
\author[a,**]{Mykola~Isaiev}
\affil[a]{Universit\'e de Lorraine, CNRS, LEMTA, 54000 Nancy, France}
\affil[*]{e-mail: liudmyla.klochko@univ-lorraine.fr}
\affil[**]{e-mail: mykola.isaiev@univ-lorraine.fr}

\begin{document}
  \maketitle
\begin{abstract}

Investigating properties of phase change materials (PCMs) is an important issue due to their extensive use in heat storage systems and thermal regulation devices. Improvement of the efficiency of such systems should be based on a better knowledge of the microscopic mechanisms governing the thermal and rheological characteristics of PCMs. This may be accomplished by the use of molecular simulations of the aforementioned quantities and their linkage with macroscale investigations.
In this work, we studied thermophysical properties of $n$-hexadecane for different temperatures regimes using molecular dynamics (MD) and carry out several experimental measurements. 
Particularly, we focused on the evaluation of various rheological and thermal properties such as thermal conductivity, $\kappa$, viscosity, $\eta$, diffusion coefficient, $D$, and heat capacities $C_p$ and $C_v$. Special attention was paid to the comparison of the results of simulations with experimental ones.
\end{abstract}



\section{Introduction}
\label{sec:intro}
Rising of the energy demand as well as challenges related to climate change urge the development of efficient energy storage technologies in accordance with the growth of renewable energies production. In this context phase change materials (PCMs) are essential to store/release heat resulting from several industrial and domestic processes. For example, fabrics for clothing that contain PCMs (such as paraffins) provide an excellent thermal regulation, and reduce the amount of sweat produced by the body. They were used at the 2004 Athens Olympics as a ``precool'' vest invented by the Australian Institute of Sport~\cite{Hu2008,Ho2011}. PCMs based systems are also inescapable to increase the efficiency of the storage of thermal energy produced from renewable sources such as solar~\cite{Chaturvedi2021} and geothermal energy~\cite{AbbasiKamazani2021}. Moreover, PCMs-based storage cells allow to collect industrial wasted energy for further re-usage, e.g. for maintaining appropriate working conditions~\cite{Yao2020}. The wide range of applications of PCMs-based thermal storage systems can be related to the significant latent heat that can be accumulated and released during the solid/liquid phase transition. Despite a large application potential, the employment of PCMs remains often limited due to the complexity of their efficient application.

The first issue is connected with a relatively small thermal conductivity ($\kappa$=0.1-0.5~W/m.K) of the most known PCMs. The latter limits theirs charging/discharging rates during solidification/melting processes. This issue can be overcome with the use of solid nanoparticles~\cite{Cheng2021, Aljaerani2022} or porous medium/solid foam~\cite{Esapour2018, Qureshi2021} to tailor thermal transport properties of the ``composite'' material. In the last example, from applied point of view, the compromise should be found between the volumetric portion of PCMs and the natural reduction of the thermal conductivity of porous matrix with porosity. Moreover, thermal contact resistance between PCM and matrix wall can significantly reduce thermal transport of PCMs-based composites~\cite{Qiu2021}. Therefore, research is currently directed on the macroscale optimization of device's configurations to achieve higher performances~\cite{Nakhchi2020}. At nano/microscale, improvements are possible through the tuning of PCMs/host matrix interactions in elaborated composite materials.

Another aspect that remains problematic in thermal energy storage systems is the  triggering of crystallization in supercooled PCMs.
Due to this feature, it is often complicated to recover the stored latent heat at the expected conditions, which causes some issues in temperature-controlled applications~\cite{Safari2017}.
Triggering the nucleation process is a possible way to address this issue~\cite{Beaupere2018}. Specifically, it can be performed by providing new nucleation sites adding new  solid/liquid interfaces.
Moreover, nucleation can be furthermore enhanced when the composite material has appropriate surfaces, for example surfaces with adapted wettability which minimizes the nucleation energy at a given temperature~\cite{Beaupere2018}.

The above-mentioned strategies for enhancing the efficiency of PCMs-based systems require an understanding of the phenomena that occur at the interface between a PCM and a nanostructured solid, because local changes in properties can be huge in systems with a large interfacial area~\cite{Isaiev2020}. Thus, the arrangement of molecules/atoms near to the interface is critical in this respect. For instance, heat and mass transfer are known to be significantly modified at the interfaces between different materials (i.e. solid and liquid ones). Noticeable changes are also reported for properties such as thermal conductivity or enthalpy of melting close to a nanoparticle~\cite{Zhao2020} or at fluid-solid interfaces in porous network~\cite{Yan2021}. 

Usually one does not know how the distribution of molecules drives the phase change process. Further research into these pathways is thus essential. On the other hand, a thorough description of the interfacial characteristics demands a correct representation of the bulk PCMs' key features. Atomistic approaches are one possible strategy to determine both structural quantities at the interface and bulk properties in such a situation. Specifically, molecular dynamics (MD) techniques are effective tools for investigating the behavior of various systems based on interactions between individual atoms. As a result, it allows us the investigation of the thermophysical~\cite{LuningPrak2021,Zhang2020,Zhao2022}, rheological~\cite{Morrow2021,Zhao2021,LuningPrak2021}, and structural~\cite{Zhao2020,Morrow2018} characteristics of PCMs and PCM-based nanocomposites. Howewer, MD simulations require  the specification of the suitable interaction potential and its subsequent parametrization. It is clear that the latter should be validated by the correlation with the experimental findings to allow further application for the reliable description of the interfacial properties.  

From the experimental viewpoint, many reviews on the properties of  PCMs are available in the literature (e.g. \citep{cabeza2015unconventional, ma12182974}). Nevertheless, the provided information is usually very specific, it often focuses on macroscopic thermal properties (melting temperature, latent heat) with partial description in terms of methods and protocols. A full thermophysical characterization is clearly complex because it involves many skills from different fields such as physics of matter, thermodynamics, mechanics, and chemistry, investigating at small scales~\cite{fernandez2015unconventional}. An additional complexity lies in the experimental measurements of PCMs properties under controlled conditions and protocols~\cite{cabeza2015unconventional}. As mentioned above, supercooling effects are strongly sensitive to experimental conditions. In some recent works~\cite{Sgreva2022, Noel2022}, we provide the background  for a complete and clearly detailed characterization procedure adapted to PCMs. This experimental background enables to measure thermophysical properties of PCMs with controlled methodologies and protocols. In this way, beyond the sole validation of one method, numerical and experimental approaches can be complementary and can enrich each other. Nevertheless, the first step to achieve consists in ensuring that the results from the two parts (numerical and experimental) are in agreement.

In this work, we propose to study the $n$-hexadecane by means of both numerical (MD) and experimental approaches. Among the numerous types of PCMs, $n$-alkanes are considered because of their chemical and thermal stability. $n$-alkanes are widely used in thermal control and energy storage applications. For example, $n$-alkanes are of interest in the design of based-on hydrocarbons superconductors~\cite{Mitsuhashi2010}, and  oleophobic surfaces~\cite{doi:10.1126/science.1148326}. Moreover, the phase transition temperature of $n$-alkanes corresponds to the human suitable diapason~\cite{Chriaa2020}. Here, we focus on the alkane which combines 16 carbon atoms, also called the hexadecane, whose chemical formula is C$_{16}$H$_{34}$.

MD simulations are based on the application of classical mechanics laws to the description of microscopic systems. This approach may forecast macroscopic thermodynamic and dynamic observable variables for various systems~\cite{tuckerman2010statistical}. Specifically, it was already demonstrated successful implementation of MD routines for the description of viscoelastic properties of PCMs~\cite{Morrow2021,Lenahan2021}. MD techniques have been used to investigate the hexadecane in several works. It has already been shown that the temperature behavior of viscosity, density, and interfacial tension estimated with MD~\cite{Lenahan2021} for hexadecane correlates well with recent data~\cite{Morrow2021,Sgreva2022}. At the same time, there is a lack of data for modelling thermophysical parameters such as heat capacity and thermal conductivity of PCMs materials, particularly those obtained with equilibrium approaches. Therefore, we focus the current study on evaluating all sets of parameters defining a thermal storage fluid with atomistic simulations. The cross-validation of the MD results with experimental data was performed to prove the validity of atomistic simulations while representing the bulk properties of PCMs~\cite{Morrow2021,LuningPrak2021,Morrow2018}. 
 Nonetheless, improving the efficiency of PCMs-based heat capacitors necessitates knowledge of the systems' thermal physical features. There are a limited number of investigations devoted to this subject in this setting. For example, it was recently demonstrated that charged nanoparticles can improve the thermal conductivity of hexadecane, according to Zhao et al~\cite{Zhao2022}. However, such investigations are partial and need additional numerical foundation. Therefore, the current article is devoted to the MD study of transport and rheological properties as well as structural ones of a PCM, the hexadecane. The numerical results are correlated with experimental ones to demonstrate the applicability of the simulations for the modeling of the bulk properties of the studied system.
 
 It should be noted that the simulations of liquid/crystalline phase transition close to the phase transition point with MD are problematic due to the small volume of the system and time of simulations. The latter leads to a low probability of the crystallization seed arising in considered volume~\cite{DaSilva2021}. Therefore, the amorphous state of the system naturally appears in the classical MD simulations, which is characteristic of supercooling effect.
The temperature range in our paper was chosen to consider the thermal physical properties of the hexadecane close to the phase transition point ($T_m$). As the first step in understanding PCMs properties, the present study focuses on the temperature range above the crystallization point ($T > T_m$). Nevertheless, we considered several temperature points below the phase transition temperature ($T < T_m$) to investigate the occurrence of some features which may be responsible for the initiation of the state transition. Therefore, we explore materials properties throughout a temperature range, including the liquid phase, amorphous state, and liquid/amorphous transition.
As the main focus, we investigated thermal (thermal conductivity, heat capacity, thermal expansion coefficient) and transport (thermal conductivity, viscosity, diffusivity) properties of PCMs, which are significant for multi-scale description of the efficiency of thermal energy storage~\cite{Sharma2022}. It is important to note that all our simulation results were obtained under equilibrium approach without setting any gradients which may perturb the liquid properties. Particularly, equilibrium MD thermal conductivity evaluation was achieved by the atomic stress components calculations with recently proposed "cendroid" form~\cite{Surblys2019, Surblys2021}. This can form the significant background for further development of the atomistic calculation approaches for engineering of the PCM-based nanocomposite systems.

The paper is organized as follows: in section~\ref{sec:Experimet}, the experimental setups are presented. In section~\ref{sec:Simulation}, MD simulation models and tools are detailed. Then, the last sections deal with calculation of thermophysical properties and their comparison to experimentally evaluated counterparts. Concluding remarks and current perspectives to this work end the paper.

\section{Materials and experimental methods}
\label{sec:Experimet}


Several batches of 99\% pure hexadecane provided by Merck (CAS number 544-76-3) were studied. 
The data sheet of the provider indicates a melting temperature of $T_m= 291.15$~K, which is in a good agreement with values given in the literature~\cite{Espeau1996,VELEZ2015383}. 
 However, due to the appearance of undercooling effects, establishing the temperature where the crystallization occurs ($T_c$) is not easy as it depends on experimental conditions. It leads to the possibility to work in a temperature range $T_c < T_m$ where hexadecane has a liquid phase~\cite{VELEZ2015383, Sgreva2022}.
 
In order to compare observations obtained with computational methods, the experimental characterization of $n$-hexadecane presented in this study was done in liquid phase, i.e. at  $T > T_m$.
Employed techniques and methods used for the characterization were similar to those detailed in \cite{Sgreva2022,Noel2022}. 
This section aims to recall these techniques as well as the main concepts  used throughout measurements. 
\subsection{Density}
\label{subsec:exp_dens}

Measurements of density, $\rho$, were carried out at ambient atmospheric pressure with a DMA~5000M Anton-Paar densimeter on a sample of a liquid hexadecane of volume $\sim$1~mL.
The obtained data  were collected every 1~K from $T=313$~K to $T=292$~K with five minutes of thermal stabilization of the sample volume during each step of lowering temperature. Due to the small volume of material employed in the analysis, such stabilizing time was sufficient in the liquid phase.
For the considered experimental temperature range, the density measurements have an accuracy of 5$\times$10$^{-6}$~g$\,$cm$^{-3}$. In addition, the coefficient of thermal expansion $\alpha_T$ at a constant pressure $p$ can be directly deduced from the temperature dependence of the density as follows:
\begin{equation}
    \label{eq:thermal_exp}
    \alpha_T = -\frac{1}{\rho \left(T\right)} \left( \frac{\partial \rho}{\partial T }\right)_p
\end{equation}

\subsection{Thermal conductivity}

The thermal conductivity, $\kappa$, of $n$-hexadecane in liquid phase was studied by using the hot tube steady state method \citep{jannot_thermal_2018}. The equipment is made of two coaxial cylinders of different diameter with a gap between them that is totally filled with $\sim 18$~mL of liquid $n$-hexadecane. 
Two thermocouples, which are located at the mid-height of the cylinders, record temperatures $T_1$ and $T_2$ at the fluid layer's boundaries, i.e. at the interfaces between the fluid and the inner and outer cylinders, respectively.
The inner cylinder is heated by Joule effect while the outer is in contact with a water recirculation maintained at controlled temperature. Once the steady state is reached, the thermal conductivity, $\kappa$, was calculated by measuring heat flux and temperature as:
\begin{equation}
\kappa(\bar{T})=\frac{\rho_e I^2 \ln(r_2/r_1)}{2 \pi^2 (r_1^2-(r_1-d)^2) (T_1-T_2)},
\end{equation}
where temperature $\bar{T}=(T_1+T_2)/2$ is the mean temperature through the PCM shell, $\rho_e$ is the specific electrical resistivity of the inner cylinder (stainless steel), $I$ is the electrical intensity, $r_1$ and $r_2$ are the radii bounding the fluid layer (with $r_1 < r_2$), $d$ is the thickness of the inner cylinder, and $(T_1-T_2)$ is the temperature variation through the fluid layer.
Measurements were done by keeping both temperatures $T_1$ and $T_2$ higher than $T_m$.

\subsection{Heat capacity}

The heat capacity, $C_p$, was estimated by differential scanning calorimetry (DSC) method. 
This technique allows the measurement of heat transfer in a small volume (less than 1 mL) of a liquid $n$-hexadecane subjected to temperature variations. 
We used a SETERAM $\mu$-DSC3 evo calorimeter  for the temperature range of $T \in [292-311]$ K.
The temperature was varied by 2 K step increments with a heating/cooling rate of 0.2 K/min between each step.
The total heat tranferred  to the PCM sample across the temperature variation between two temperature steps is correlated with the heat capacity. The total heat in our studies is calculated by integrating the heat power transferred over time: from the start of a temperature step until there is no longer heat exchange between the PCM sample and the DSC.
As a result, each temperature step was kept constant for a long enough period of time (about 1 hours) to ensure that the heat flow was fully dissipated.

\subsection{Viscosity}
\label{exp:visc}
The rheological behaviour of liquid $n$-hexadecane was investigated with an AR-G2 rheometer (TA Instruments), by using both a plate-plate and a cone-plate geometry.

We used a 60 mm diameter plate with a fixed 1000~$\mu$m gap between the two plates in the plate-plate geometry. In the cone-plate geometry the cone angle  was 2$^{\circ}$, the plate diameter was 60 mm, and the gap was fixed at 70 $\mu$m. Such a large diameter was chosen for the cone and plate because of the low viscosity of $n$-hexadecane in a liquid phase. Both geometries produced similar results.  
Rheometry was done under isothermal conditions via a Peltier plate within a range of 289 K to 313 K. 
In the liquid phase, we observed a Newtonian behavior for hexadecane.
Creep tests were thus performed at each temperature by applying a shear stress of 0.1 Pa.
Tests lasted around two minutes, assuring a steady state response and a viscosity plateau.
The resulting viscosity at each temperature corresponds to the mean of measurements taken over this plateau. 

\subsection{Diffusion coefficient}

Nuclear magnetic resonance (NMR) was used to quantify the diffusion coefficient (which indicates the mass diffusivity of a liquid hexadecane) at various temperatures ranging from 293 K to 310 K. A split air flow, one from below and one from the side of the sample, has been used to impose the temperature on the sample (600 $\mu$L volume). Isothermal conditions were ensured by waiting 30 minutes for the temperature to equilibrate before the acquisition.

Spectra were acquired by a Bruker Avance III HD 300~MHz spectrometer equipped with a Bruker BBO probe. 
We used a stimulated echo technique with bipolar gradients \cite{Tanner1970}, a diffusion gradient duration of $\delta$=2 ms and a diffusion time interval $\Delta$=200 ms.
We acquired several spectra at constant temperature with a gradient power $g_p$ ranging from 9.63 to 471.87 mT$\,$m$^{-1}$. Eight scans were acquired for each spectrum.

\begin{figure}[h!]
        \centering
         \includegraphics[scale=0.85]{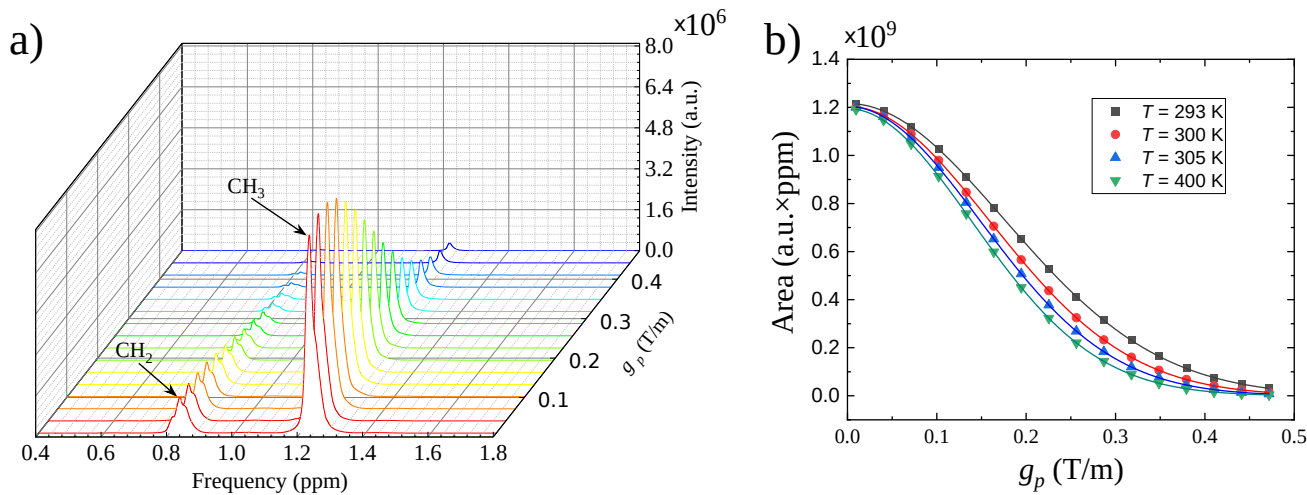}
         \caption{a) Spectra of $n$-hexadecane at $T=$293 K for different power gradients, and b) spectra area, $A$, as a function of power gradient values at different temperatures. Symbols represent the experimental points and solid lines are their fit by following an equation (\ref{eq:diff_coef}).}
         \label{fig:spectra_multiple}
\end{figure}

The collected spectra for different gradients are presented in Fig.~\ref{fig:spectra_multiple}a.
Fig.~\ref{fig:spectra_multiple}b illustrates the dependency of the spectra area vs. the gradient at various temperatures.

At a fixed temperature, with the gradient rise, the decrease of intensity of the spectra (see Fig.~\ref{fig:spectra_multiple}b) can be fitted by the Stejskal-Tanner equation \cite{stejskal_spin_1965} to obtain the diffusion coefficient $D$:

\begin{equation}
\label{eq:diff_coef}
A=A_0\times \exp(-\gamma^2 g_p^2 \delta^2 D \Delta ),
\end{equation}
where $A$ is the spectrum area, $A_0$ is the area of a spectrum with a null gradient intensity,  and $\gamma$ is the gyromagnetic ratio of the proton ($\gamma$=42.576 MHz$\,$T$^{-1}$).

\section{Molecular Dynamics simulations}
\label{sec:Simulation}

\subsection{Atomistic model for organic compounds}
\label{subsec:MD-model}

To investigate heat transfer in organic systems one can chose ``all-atom'' molecular dynamics (MD) algorithm (i.e., each atom of the studied molecule is considered as an individual element in opposite to the coarse-grain method). The latter is used to determine the thermodynamic and rheologic properties of hexadecane. In the following sections we discuss the simulation protocol and the theoretical background that were used to obtain and treat simulation data. 

\begin{figure}[h!]
\begin{center}
\includegraphics[scale=0.25]{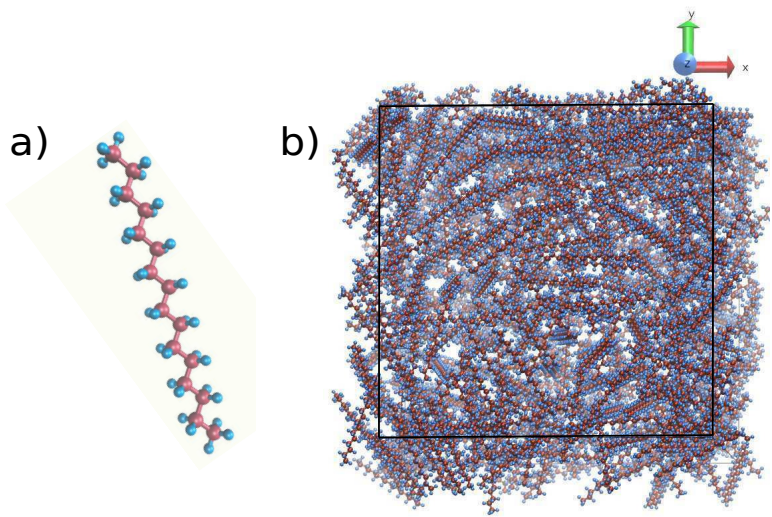}
\end{center}
\caption{\label{fig:initial} Snapshots of: a) one molecule of $n$-hexadecane,  and b) the initial system of $N=500$ molecules of $n$-hexadecane.}
\end{figure}

\subsection{Simulation protocol}
\label{subsec:MD-protocol}

In order to construct a realistic system representative of hexadecane PCMs with randomly distributed molecules we average all our calculations over $m$=7 distinct initial systems. Each of them is built with $N=500$ molecules of hexadecane by using Packmol~\cite{PACKMOL} software. Molecules are randomly set inside a cubic box with periodic boundary conditions. A system, in the initial configuration, was visualized with VMD~\cite{HUMP96,STON1998} and is represented in Fig.~\ref{fig:initial}  and setting parameters in Table.~\ref{tab:sys}. 


\begin{table}[h!]
\begin{center}
\begin{tabular}{ |c|c|c|c|c| } 
\hline
 \bf{N molecules} & \bf{$L_x$} (\AA) & \bf{$L_y$}  (\AA) & \bf{$L_z$}  (\AA) \\
\hline
 500   &  67.56   &  67.56 & 67.56\\
\hline
\end{tabular}
\caption{Initial parameters for simulation \label{tab:sys}}
\end{center}
\end{table}

Here we partly follow the protocol proposed by Morrow et al.~\cite{Morrow2021}. We chose the initial volume to match the experimental densities at different temperatures. The latter are shown in Fig.~\ref{fig:densities} (see experimental red dots) for an extended range of temperatures: $T \in [310,276]$ K with steps of 2 K. In order to speed up the packing we increased the box size of simulation cell by 5~\AA~in each dimension \cite{Morrow2021}.

\begin{figure}[h!]
\begin{center}
\includegraphics[scale=0.25]{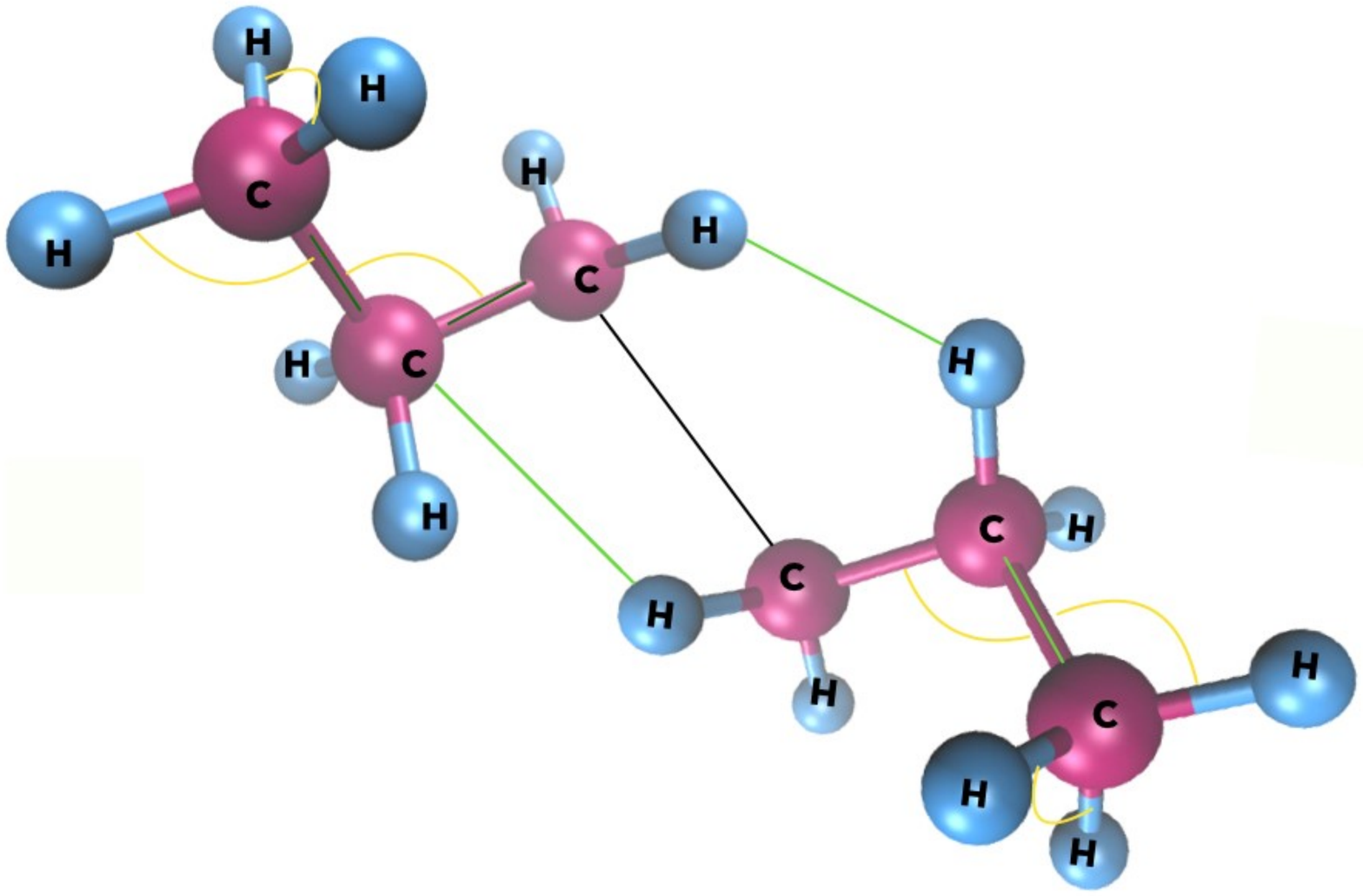}
\end{center}
\caption{\label{fig:interaction} Schematic view of the interaction between constitutive atoms of hexadecane. The black line corresponds to the non-bonded interaction with parameters shown in Tab.~\ref{tab:nb}, the green linens relate to Tab.~\ref{tab:b}, and the yellow to Tab.~\ref{tab:a}  respectively.}
\end{figure}

Molecular dynamics simulations are performed with LAMMPS~\cite{LAMMPS} software with time-step $\delta t$=2~fs. As the choice of the force field is a crucial issue, within this work all-atom optimized L-OPLS~\cite{LOPLS} force field (see Tables.~\ref{tab:nb}--~\ref{tab:dihedral} for more detail)  was used due to its realistic approximations compared to experiments~\cite{Morrow2021, Moultos2016, Papavasileiou2019, Guido2019, Siu2012} (see also secs.~\ref{sec:Experimet} and~\ref{sec:conclusions}). The example of different types of interactions, such as bonded, non-bonded, and angle interactions, are presented in Fig.~\ref{fig:interaction}. In this work the truncation of non-bonded interaction at $r_{\text{cut}}$=17~\AA~was used. Carbon-hydrogen bond distances were constrained by applying the SHAKE~\cite{RYCKAERT1977327} algorithm. This algorithm applies bond and angle restrictions to given bonds and angles in the simulations, i.e. each timestep specified angles and bonds are reset to their equilibrium values.   


\begin{table}[h!]
\begin{center}
\begin{tabular}{ |c|c|c|c|c| } 
\hline
 \bf{Atom} & \bf{Partial charge} ($e$) & $\sigma$  (\AA) & $\varepsilon$ (kcal/mol) \\
\hline
 C (hexadecane CH$_3$)   & -0.222    &  3.50 & 0.06600\\
 C (hexadecane CH$_2$)   & -0.148    &  3.50 & 0.06600\\
 H (hexadecane CH$_3$)   & 0.074    &  2.50 & 0.03000\\
 H (hexadecane CH$_2$)   & 0.074    &  2.50 & 0.02629\\
\hline
\end{tabular}
\caption{Non-bonded parameters for L-OPLS force field~\cite{Siu2012} \label{tab:nb}}
\end{center}
\end{table}

\begin{table}[h!]
\begin{center}
\begin{tabular}{ |c|c|c|c|c| } 
\hline
 \bf{Bond} & $r_{\text{eq}}$  (\AA) & $K_b$ (kcal/mol) \\
\hline
 CT-CT   & 1.529    &  268\\
 CT-H   & 1.090    &  340\\
\hline
\end{tabular}
\caption{Bonded parameters for L-OPLS force field~\cite{Siu2012} \label{tab:b}}
\end{center}
\end{table}

\begin{table}[h!]
\begin{center}
\begin{tabular}{ |c|c|c|c|c| } 
\hline
 \bf{Angle} & $\theta_{\text{eq}}$  (degrees) & $K_{\theta}$ (kcal/mol) \\
\hline
 CT-CT-CT   & 112.7 &  58.35\\
 CT-CT-H   & 110.7  &  37.50\\
 H-CT-H   & 107.8  &  33.00\\
\hline
\end{tabular}
\caption{Angle parameters for L-OPLS force field~\cite{Siu2012} \label{tab:a}}
\end{center}
\end{table}

\begin{table}[h!]
\begin{center}
\begin{tabular}{ |c|c|c|c|c| } 
\hline
 \bf{Dihedral} & $V_1$ & $V_2$ & $V_3$ \\
\hline
  CT-CT-CT-CT   & 0.6447    &  -0.2143 & 0.1782\\
\hline
\end{tabular}
\caption{Dihedral Fourier parameters (kcal/mol) for L-OPLS force field~\cite{Siu2012} \label{tab:dihedral}}
\end{center}
\end{table}

\subsection{Liquid/amorphous state, a molecular investigation} 

The first part of this work is devoted to the characterization of molecular organization of the studied $n$-hexadecane samples considered in MD simulations used to evaluate thermophysical properties. As mentioned above, crystalline structure of $n$-hexadecane~\cite{Zhao2022} are scarcely achievable with the simulation parameters used here. Instead, we suggest that amorphous state, characterized by amorphous molecules organization is more probably reached when the calculations are performed below the melting temperature $T_m$, i.e. in the undercooling region. The purpose of this section is to clarify this point through the use of radial distribution function.

As the solidification process is not changing the initial structure of sample (liquid or solution) it was chosen to calculate radial distribution function (rdf), $g(r)$, of $n$-haxedecane at different $T$'s regimes. The main aim of performing such calculations is to observe the possible reorganization of molecule constitutive atoms while the system cross the phase transition temperature.  In other words, if patterns of rdf remain identical for "liquid" ($T > T_m$) and "solid" ($T < T_m$) characteristic temperature of the system, it confirms that the latter reaches an amorphous state while the temperature is cool down in MD simulation and goes bellow $T_m$.
Radial distribution function, $g(r)$, can be calculated as: 
\begin{equation}
\label{eq:rdf}
    g(r) = \frac{1}{\rho}\left< \sum_{i \neq 0} \delta (\boldsymbol{r} - \boldsymbol{r}_i) \right>
\end{equation}
where $\delta (\boldsymbol{r})$ is the delta function at position $\boldsymbol{r}$. 

\begin{figure}[h!]
\begin{center}
\includegraphics[scale=0.32]{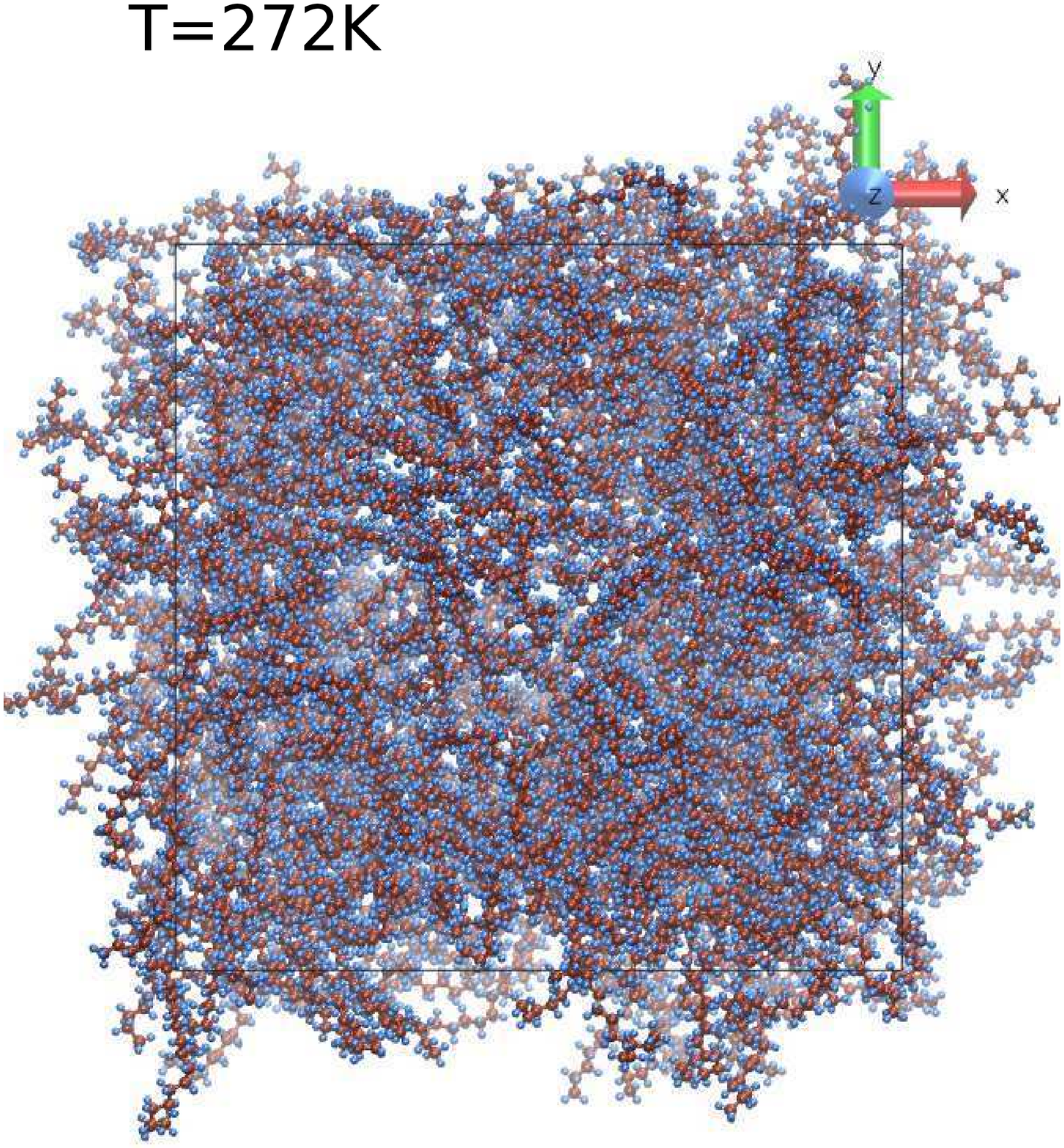}
\includegraphics[scale=0.32]{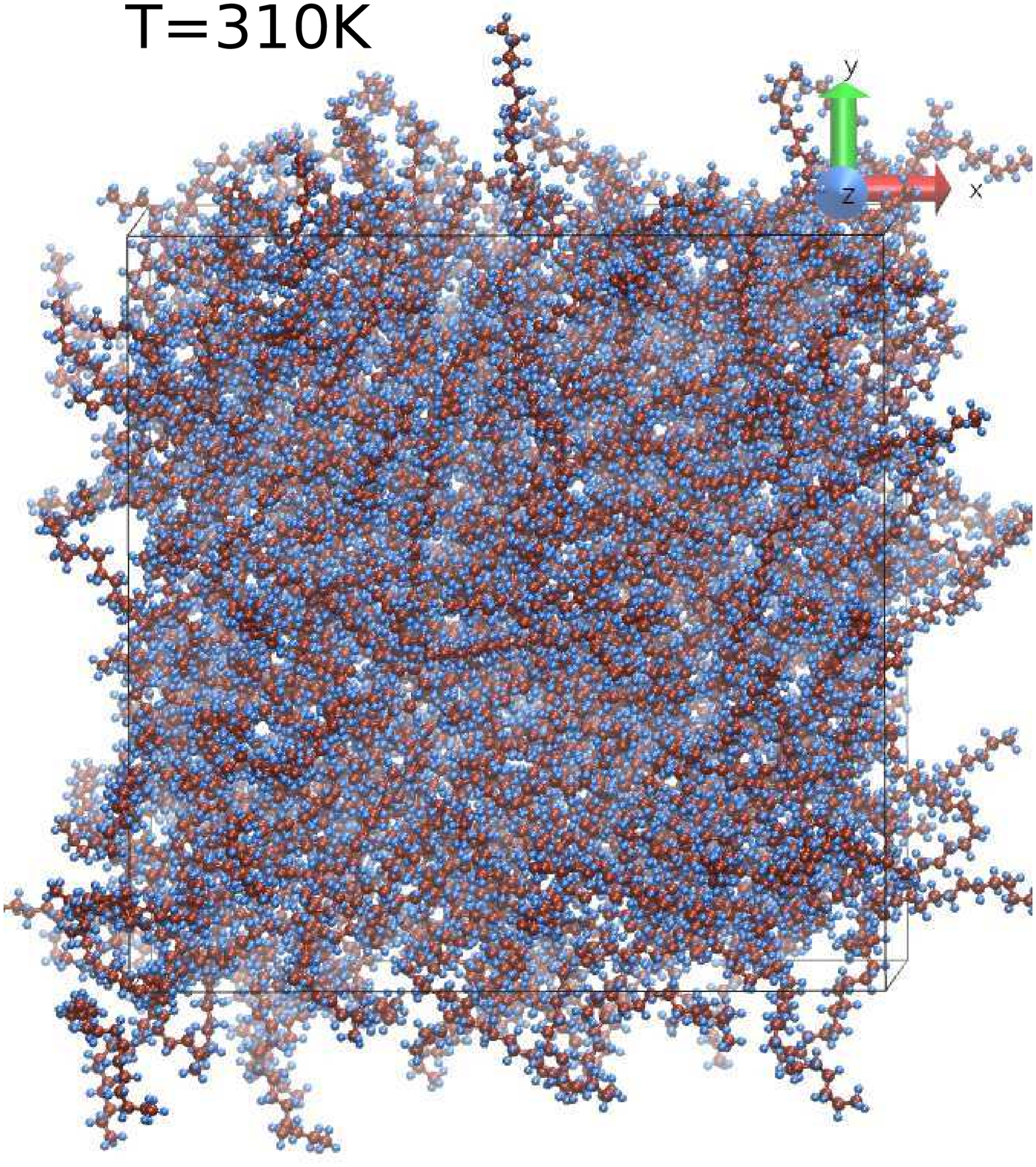}
\includegraphics[scale=0.35]{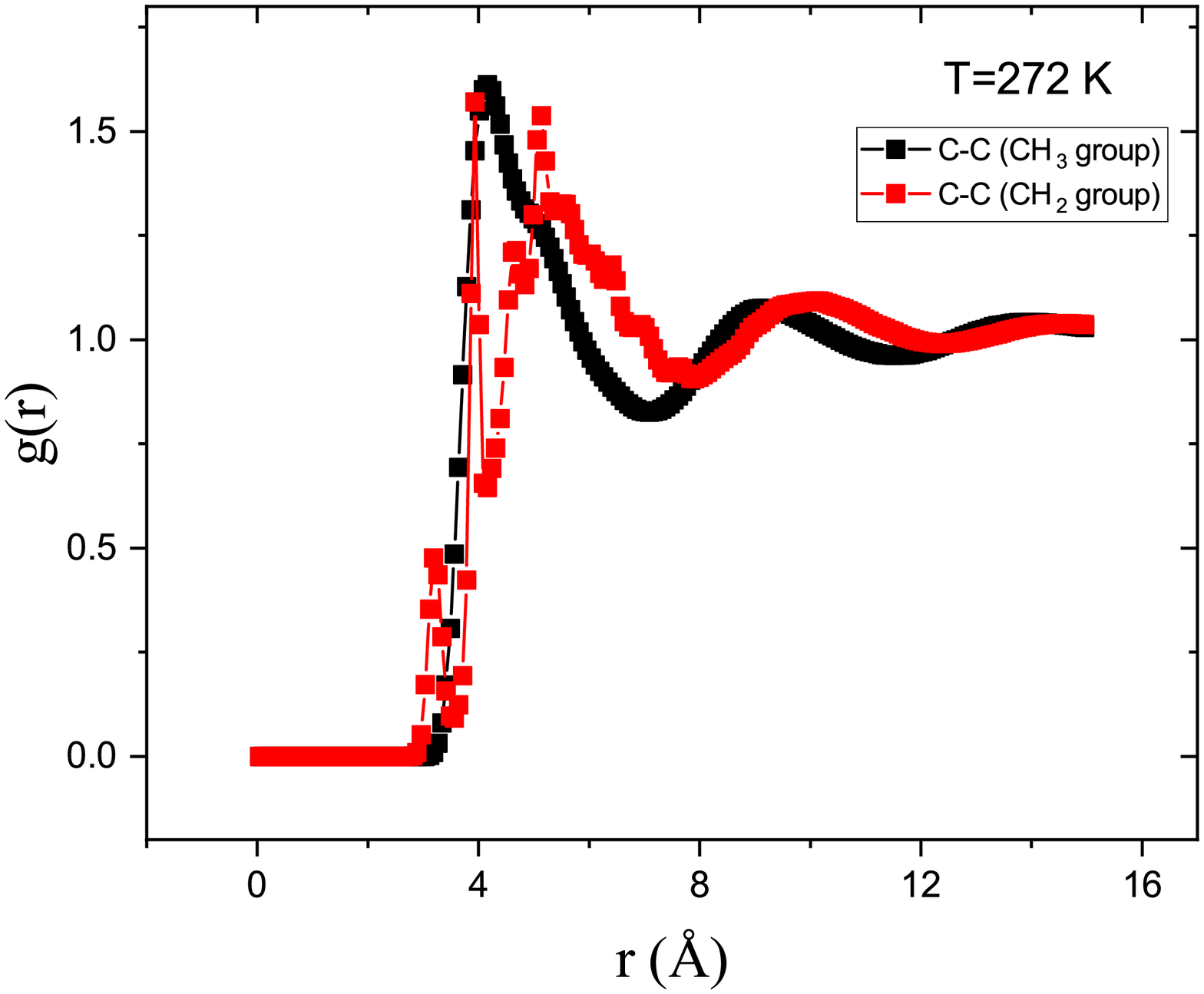}
\includegraphics[scale=0.35]{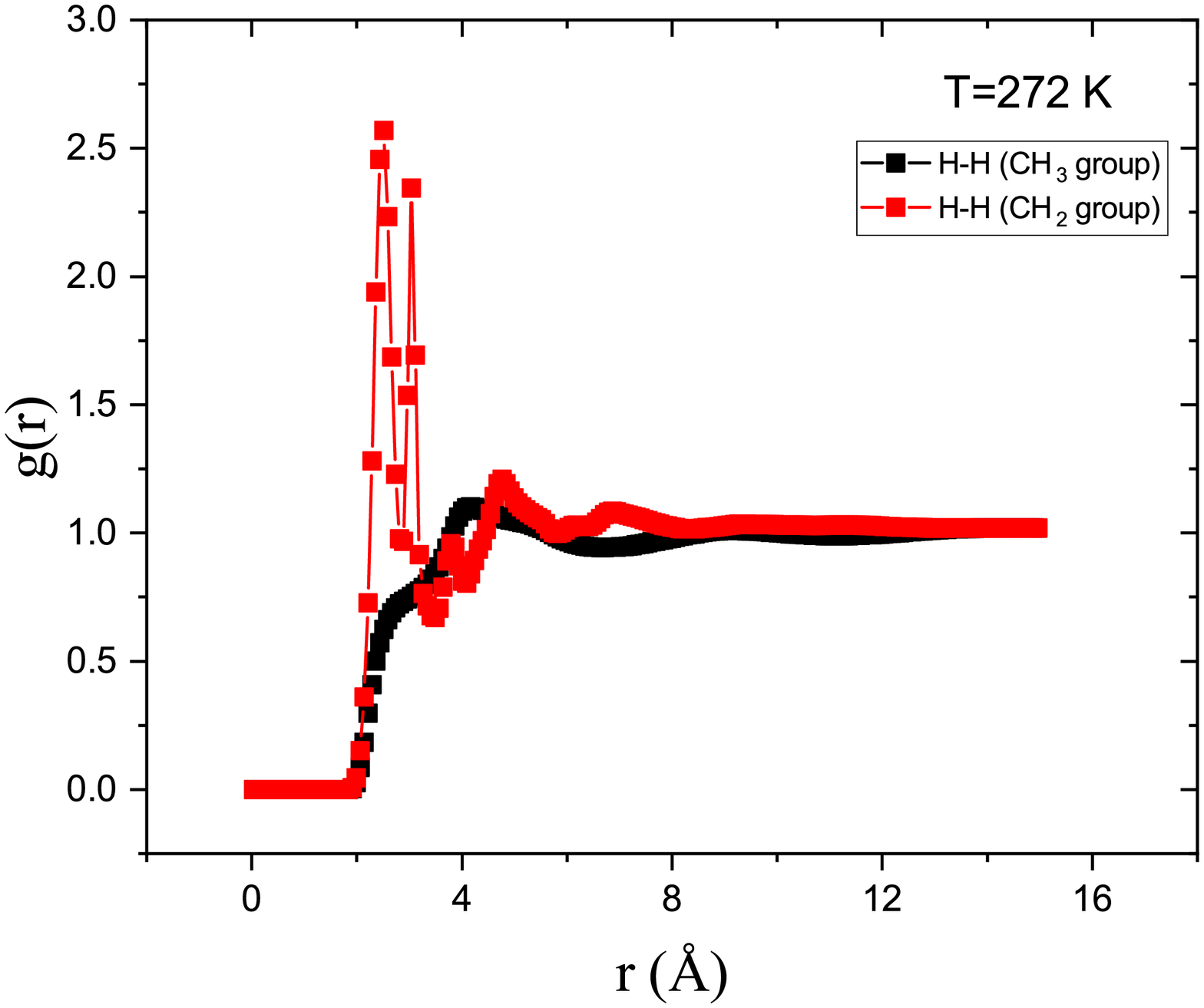}
\includegraphics[scale=0.35]{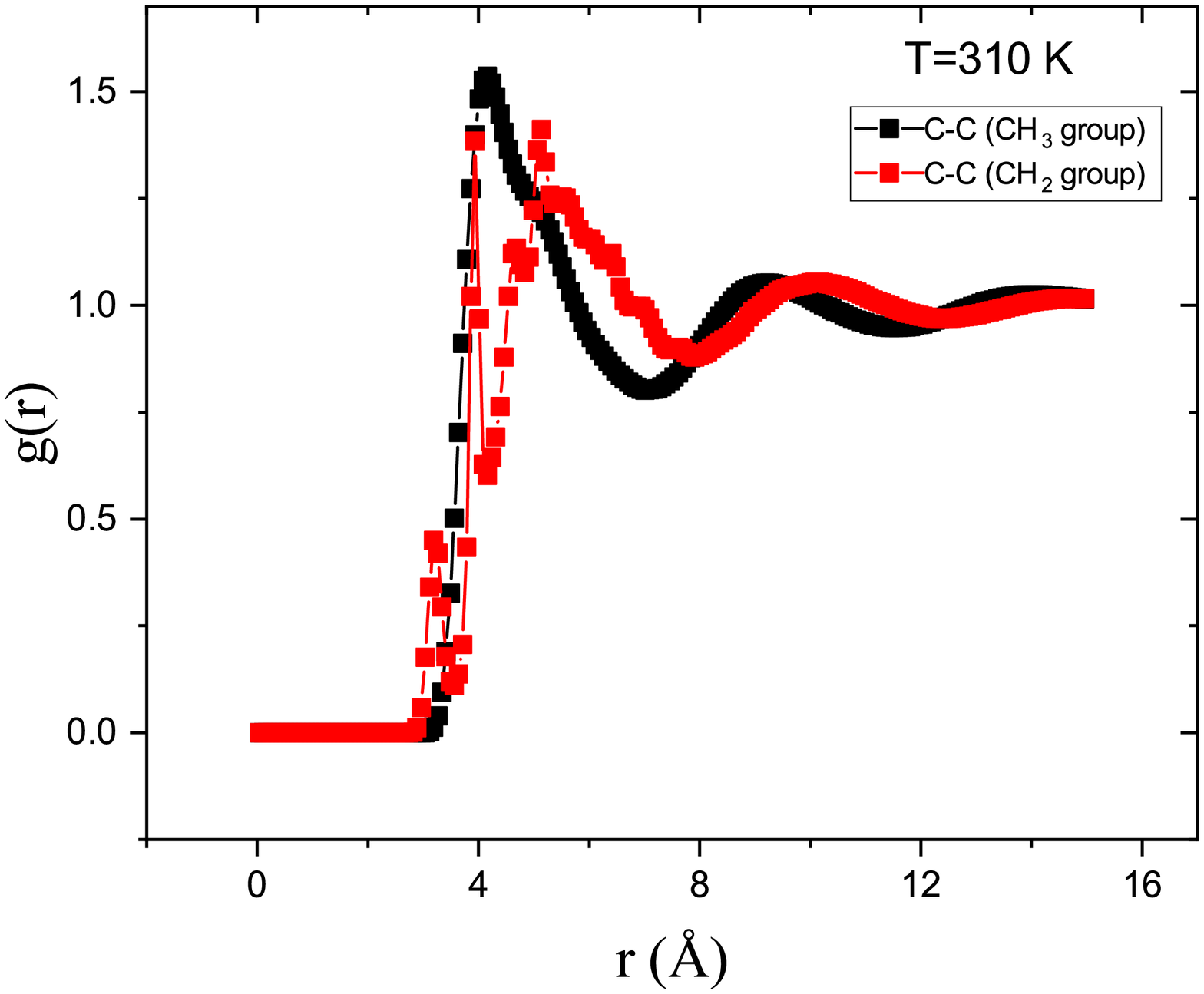}
\includegraphics[scale=0.35]{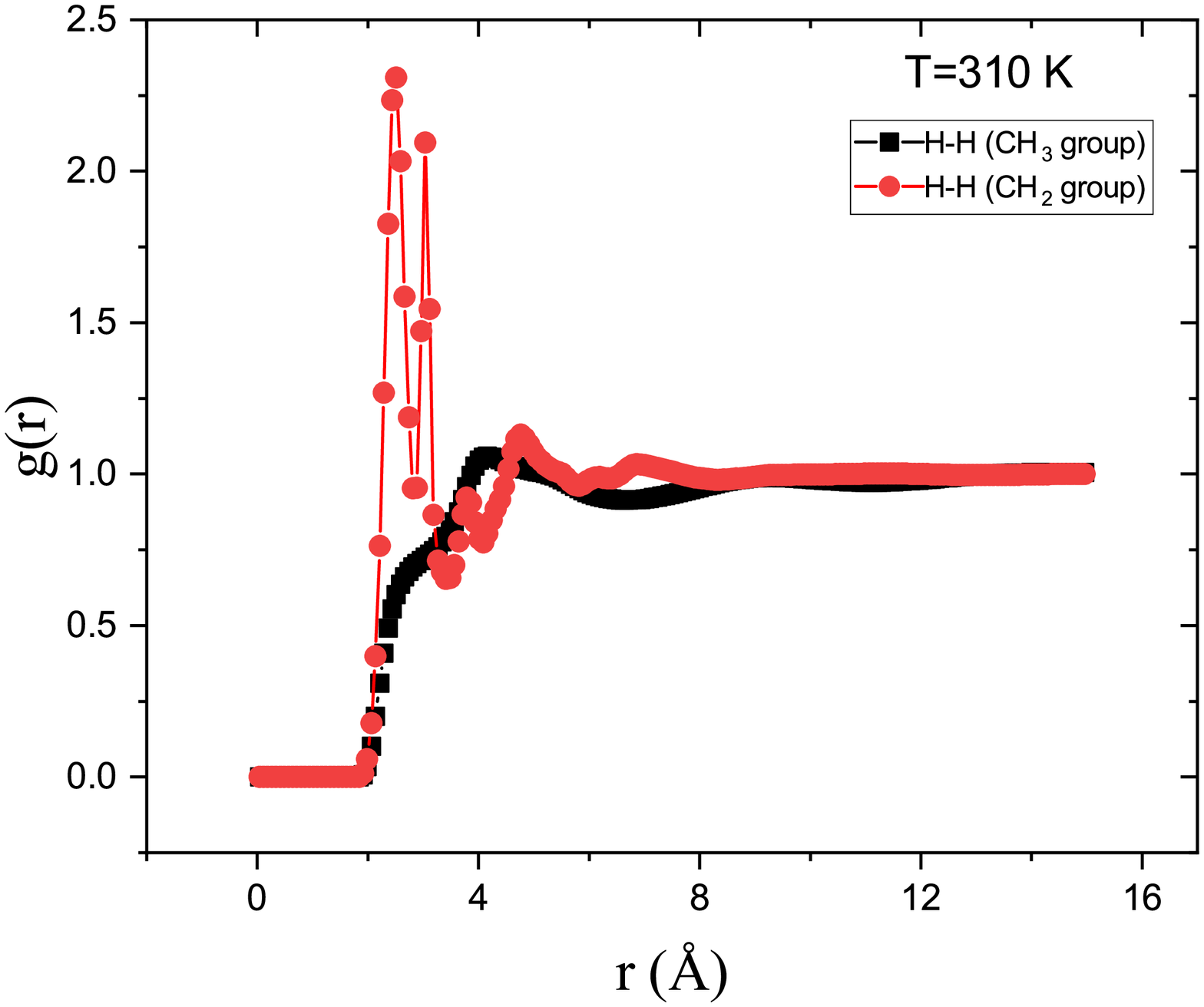}
\end{center}
\caption{\label{fig:rdf_272-310} Snapshots $T=272$~K and $T=310$~, and radial distribution function at $T=272$~K (top figures) and $T=310$~K (bottom figures).}
\end{figure}

We calculated the radial distribution function between different types of atoms: i) carbon-carbon atoms in CH$_3$ and CH$_2$ groups; ii)  hydrogen-hydrogen atoms in CH$_3$ and CH$_2$ groups.  
Results for $T$=310~K and $T$=272~K are plotted in Fig.~\ref{fig:rdf_272-310}. As it can be clearly seen, rdf peaks remain at the same location, there is only a slight change in their amplitude when $T$ varies. This undoubtedly confirms that no phase transition with a noticeable change of molecules organization occurs. In the achieved simulation our systems for the whole temperature range remain in amorphous state.

\subsection{Thermal volume expansion of the $n$-hexadecane via MD}
\label{subsec:MD-evaluation}

Simulations start with energy minimization of the system by updating atom coordinates continuously with the use of ``minimize'' command in LAMMPS (for more detail see \url{https://docs.lammps.org/minimize.html}). Then, in a second stage,  the simulations are switched in $NVT$ ensemble (where $N$ is the total number of molecules, $V$ and $T$ are the volume and the temperature of the system)  for a typical duration of $t$=245~ps applying Nos{\'e}-Hoover thermostat to control temperature~\cite{Nose:JCP1984} and to adjust the chosen volume to the system. Then, in a third stage, in order to reach the desired pressure, $p$=0.1~MPa, we perform $NPT$ equilibration run for $t$=4~ns.
During a remaining simulation time $t$=2~ns, information about density, $\rho$, was registered every $t=20$ fs. 

\begin{figure}[h!]
\begin{center}
\includegraphics[scale=1]{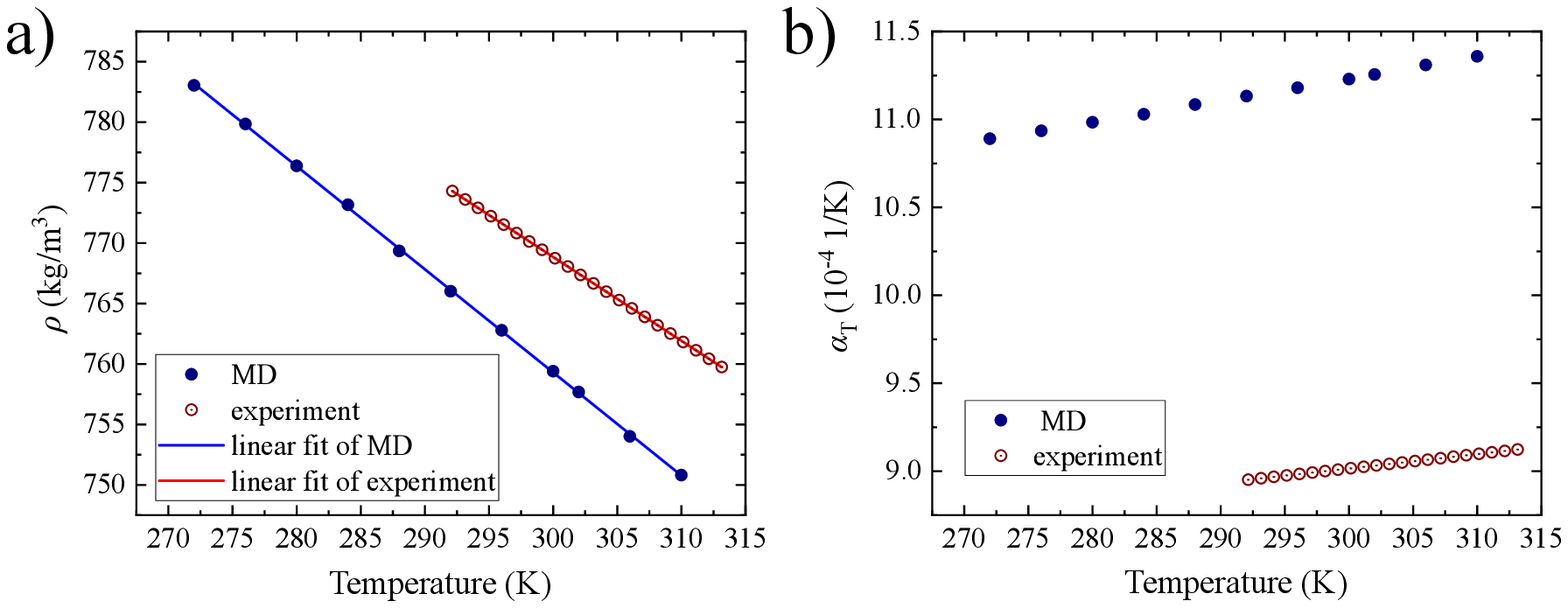}
\end{center}
\caption{\label{fig:densities} Temperature dependence of hexadecane: a) density, $\rho$, and b) thermal expansion coefficient, $\alpha_T$,  obtained with MD simulations (blue dots) and  experimentally (maroon dots)~\cite{Sgreva2022}. Error bars are not indicated due to their insignificance.}
\end{figure}

In Fig.~\ref{fig:densities} the averaged values of hexadecane density obtained by the above detailed simulation procedure are plotted with experimental data obtained by following protocol from subsec.~\ref{subsec:exp_dens}. As it can be seen, the estimated density from MD is in a great agreement with the actual experimental data, which decreases linearly with increasing temperature; our findings are similar with earlier investigations~\cite{Lal2000}. In this latter study was experimentally founded the typical values for density of $n$-hexadecane in liquid phase ($T=298.15$ K) $\rho=770.6$ kg m$^{-3}$ that is close to that we obtained within our research. In the considered temperature range, the calculated error is lower than 5\%. 
 
Knowing the temperature dependence of $n$-hexadecane's density, one can straightforwardly obtain the thermal volume expansion coefficient $\alpha_T$ (see eq.~\ref{eq:thermal_exp}).
Using the values of $\rho$ given in Fig.~\ref{fig:densities} one can obtain $\alpha_T(T)$. The calculated dependence is shown in Fig.~\ref{fig:densities} for both simulation and experimental data~\cite{Sgreva2022, Noel2022}.  These first results on structural properties ($\rho$ and $\alpha_T$) of the modelled hexadecane make us confident about the relevance of the chosen atomic model. In the following section, thermal properties will be thus investigated.

\section{Calculation of thermophysical properties}
\label{sec:CTP}
 
For the calculation of thermal conductivity and viscosity of the system, we change slightly the above presented equilibration procedures. Specifically, additional equilibration run in canonical ensemble ($NVT$) for $t$ =4~ns was performed due to obtain as much equilibrated system as possible. Then, we ensure a long relaxation stage, the system remains in $NVT$ ensemble for $t$=4~ns instead of 2~ns. Relevant information is saved each $t$=320~ps for viscosity, $\mu$,  each $t$=80~ps for thermal conductivity $\kappa$, and each $t$=2~ps for total energy of the system, $E$, entalphy, $H$, positions, and mean-square displacement.

Achieving such calculations that take into account all the interactions is costly in terms of simulation time and memory resources. We thus perform the above procedure with $m$=5 independent configurations for each temperature point to achieve the optimum and desired results.

\subsection{Viscosity and dynamical properties}

We used the Green-Kubo method with ensemble average of the auto-correlation of the shear stress tensor to calculate viscosity as: 
\begin{equation}
\label{eq:GK_visc}
\mu = \frac{V}{k_BT}\int_0^{\infty} \left<  \tau_{\alpha \beta}(t) \tau_{\alpha \beta}(0)\right>dt
\end{equation}
where $V$ is volume of the system at temperature $T$, $k_B$ is Boltzman's constant, and $\tau_{\alpha \beta}$ are the non-diagonal elements of the stress tensor ($\alpha\beta$=$xy$, $yz$ or $xz$). 

\begin{figure}[!ht]
\begin{center}
\includegraphics[scale=1]{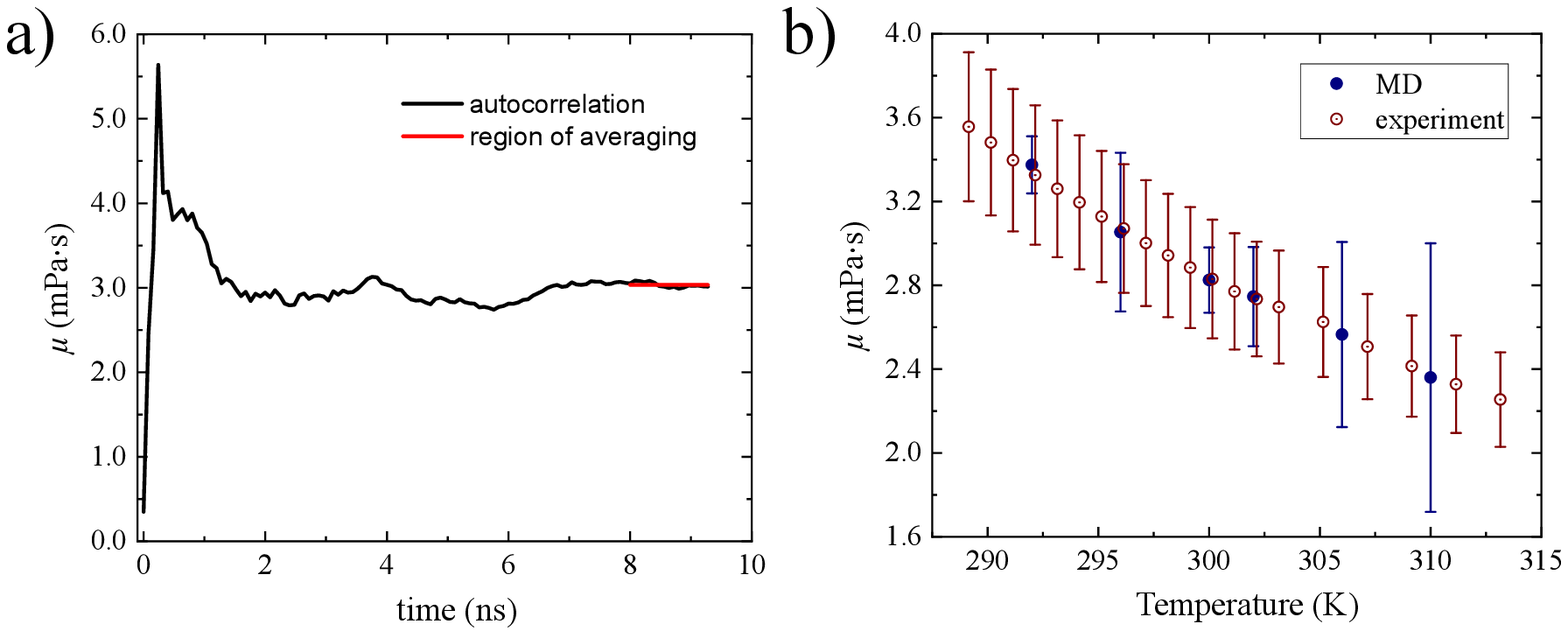}
\end{center}
\caption{\label{fig:visosity} a) Autocorrelation viscosity function obtained for $T=296$ K. The red line corresponds to the time interval that was used to calculate the mean value shown in Fig.~\ref{fig:visosity} b) and b) dynamic viscosity as a function of temperature.}
\end{figure}

Knowing that Eq.~\ref{eq:GK_visc} can only be used for ``well equilibrated'' systems (i.e., for simulation time much higher than the relaxation time of the system) with the use of LAMMPS one has to be sure that steady state is reached. An example of the averaging interval for $\mu(t)$ is shown in Fig.~\ref{fig:visosity} a. In this work, careful evaluation of the ``plateau''-like interval was achieved to ensure accurate evaluation of the dynamic viscosity  $\mu(t)$ at each desired temperatures. Furthermore, dynamic viscosity of $n$-hexadecane was measured using plate-plate rheometer controled with Peltier cooling as it was detailed in sec~\ref{exp:visc}. The comparison between numerical and experimental data is given in Fig.~\ref{fig:visosity} b. The agreement is very good on the whole temperature range.
Moreover, in literature~\cite{Lal2000} (at $T=298.15$ K) measured viscosity is $\mu=3.039$ mPa$\cdot$s. This demonstrates the reliability of the proposed techniques.  
\subsection{Thermal conductivity}
 \label{subsec:thermal_conductivity}

Thermal conductivity can be also calculated with EMD according to Green-Kubo formalism through the integration of heat flux autocorrelation, as follows: 
\begin{equation}
    \label{eq:GK_therma}
    \kappa = \frac{V}{k_BT^2} \int_{0}^{\infty}\left< \boldsymbol{J}\left(t\right)\boldsymbol{J}\left(0\right)\right> dt
\end{equation}
where  $\boldsymbol{J}(t)$ is the heat flux vector at time $t$, that can be obtained from:
 \begin{equation}
    \label{eq:heat_flux}
    \boldsymbol{J} = \frac{1}{V}\Big[\sum_i\boldsymbol{e}_i\boldsymbol{v}_i - \frac{1}{2}\sum_{i<j}\boldsymbol{F}_{ij}\cdot(\boldsymbol{v}_i+\boldsymbol{v}_j)\boldsymbol{r}_{ij}\Big]
\end{equation}
where $\boldsymbol{e}_i$ is the total per-atom energy, $\boldsymbol{F}_{ij}$ is the force between atoms $i$ and $j$, $\boldsymbol{v}_i$ and $\boldsymbol{v}_j$ are their velocities. Finally, $\boldsymbol{r}_{ij}$ is the distance between atoms $i$ and $j$. 
 
\begin{figure}[!ht]
\begin{center}
\includegraphics[scale=0.35]{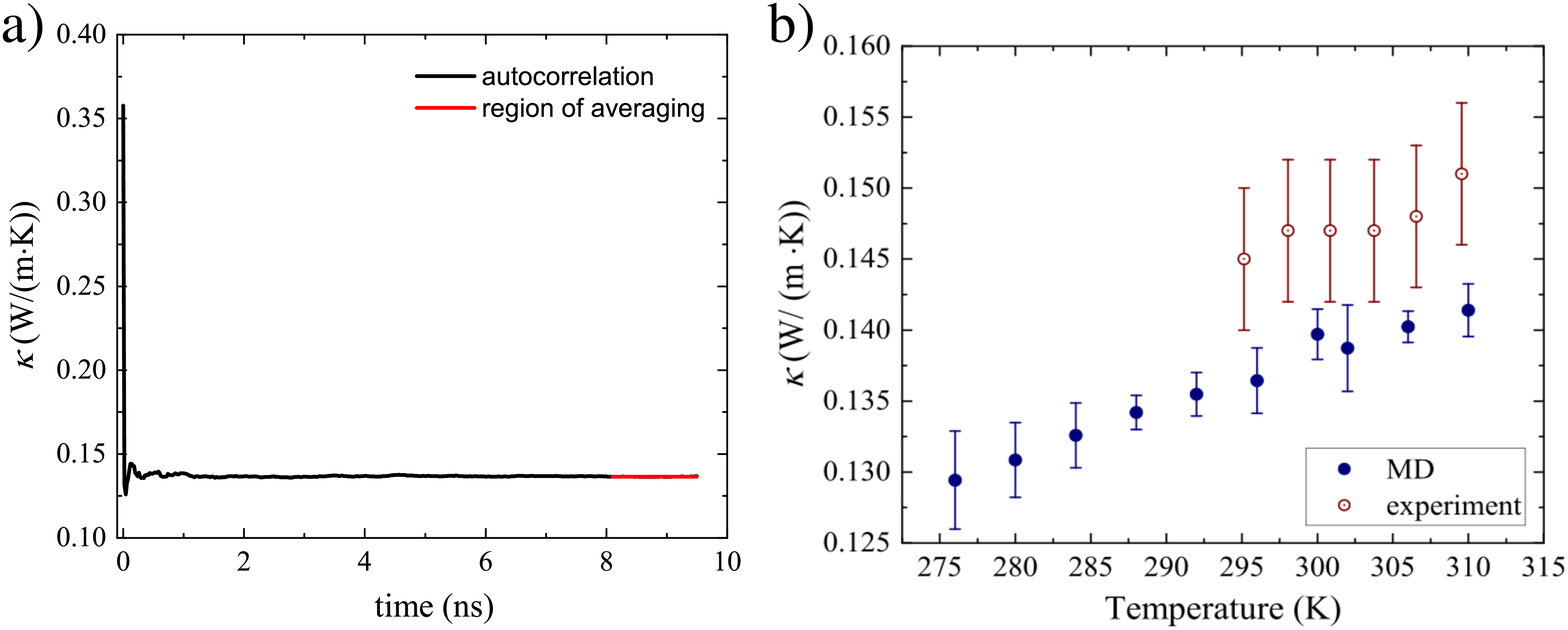}
\end{center}
\caption{\label{fig:conductivity_comp} a) Autocorrelation thermal conductivity function obtained for $T=296$ K. The red line corresponds to the average interval that was used to calculate the mean value shown in Fig.~\ref{fig:conductivity_comp} b) and b) Temperature dependence of thermal conductivity, $\kappa$,  obtained by two different approaches: MD (blue dotes) and experiments (maroon dotes).}
\end{figure}

The solution of Eq.~\ref{eq:GK_therma} is well-known. In our calculations the usage of ``centroid/stress/atom'' option integrated in LAMMPS simulation package~\cite{doi:10.1021/acs.jctc.9b00252,PhysRevE.99.051301} was considered. Both numerical and experimental data for thermal conductivity are plotted as a function of temperature $T$ in Fig.~\ref{fig:conductivity_comp} b. First, let us analyze a reduced temperature range (i.e. [295-310~K]) where comparisons can be made between both approaches. In this range temperature is above liquid-solid phase transition of $n$-hexadecane. Here, in liquid state, thermal conductivity evaluation by both approaches matches. We can observe that thermal conductivity remains constant around 0.14~W m$^{-1}$K$^{-1}$ in the studied temperature range.

In the frame of this study, detailed simulation by MD of liquid-solid phase transition features was not done as the crystalline-like structure of the solid $n$-hexadecane cannot be recovered from the chosen initial liquid state. As a result, the majority of the comparisons were made using our MD simulations and our own experimental data, which corresponds to the liquid phase behavior above $T$=292 K (i.e. above the melting temperature $T_m$). However, numerical simulations allow us to achieve ``excursions'' into lower temperature regimes, below the state transition limit, and thus to extract some insights about physical properties. We perform such calculations where the considered $n$-hexadecane sample is in an ``amorphous  state'' as was discussed previously. Such computations were carried out for the thermal conductivity decreasing $T$ till 276 K, a temperature much below the melting point $T_m$. The findings of our MD simulations of thermal conductivity are presented in  Figs.~\ref{fig:conductivity_comp}. As it can be seen, continuous increase of $\kappa$ is observed from 276~K to 310~K, except temperature region $T \in [296, 302]$ K where a small peak can be observed. Such weak non-monotonic dependence of thermal conductivity on temperature was previously observed for amorphous silicon~\cite{Isaeva2019}.

This peak-like behaviour at given temperature range arises because of the specific heat jump (see ~\cite{DosSantos2013}, and also observed in Fig.~\ref{fig:CP}), and it can be explained from the point of view that hexadecane system is facing liquid-amorphous state transition. The values of $T$'s within observed range are above to the $T_m$ obtained within experiment. Yet, they remain much higher than the one measured in other experimental studies (see right plot of Fig.~2 in ref.~\cite{inproceedings} for instance). Considering our MD simulation outputs for thermal conductivity, we can assume that liquid-amourphous state transition should occur in the range of $T_g \in \left(296,302\right)$ K.

\subsection{Constant pressure and volume heat capacities}

Heat capacities at constant pressure (isobar), $C_p$, and constant volume (isochor), $C_v$ have been also investigated by means of atomic scale simulations.
Molar isochore heat capacity at  temperature $T$,  can be related to the system energy fluctuations, in the canonical ensemble it can be calculated as~\cite{doi:10.1063/5.0046697}: 
 \begin{equation}
 \label{eq:heat_capacity_general}
    C_v (T) = \frac{\sigma^2_E}{k_BT^2},
 \end{equation}
where $\sigma^2_E$ is the variance of the total energy.  The variance of the total energy, $\sigma^2_E$, was obtained from the simulation data that were kept after the final $NVT$ production run  (see sec.~\ref{subsec:MD-protocol}) with the use of refs.~\cite{George2021A,George2021B} approaches. As it can be seen from the sec.~\ref{subsec:MD-protocol}, after the production run the set of energies, $E(t)$ as a function of time, $t$, was obtained for each configuration. With the use of trivial relation to obtain variance, we calculated $\sigma^2_E$ at each temperature in a row:
\begin{equation*}
    \sigma^2_E = \frac{1}{n-1}\sum^n \left(E_i-\overline{E} \right)^2
\end{equation*}
where $n$ is the number of points in a set, and overbar corresponds to the mean value. 

\begin{figure}[!ht]
\begin{center}
\includegraphics[scale=0.5]{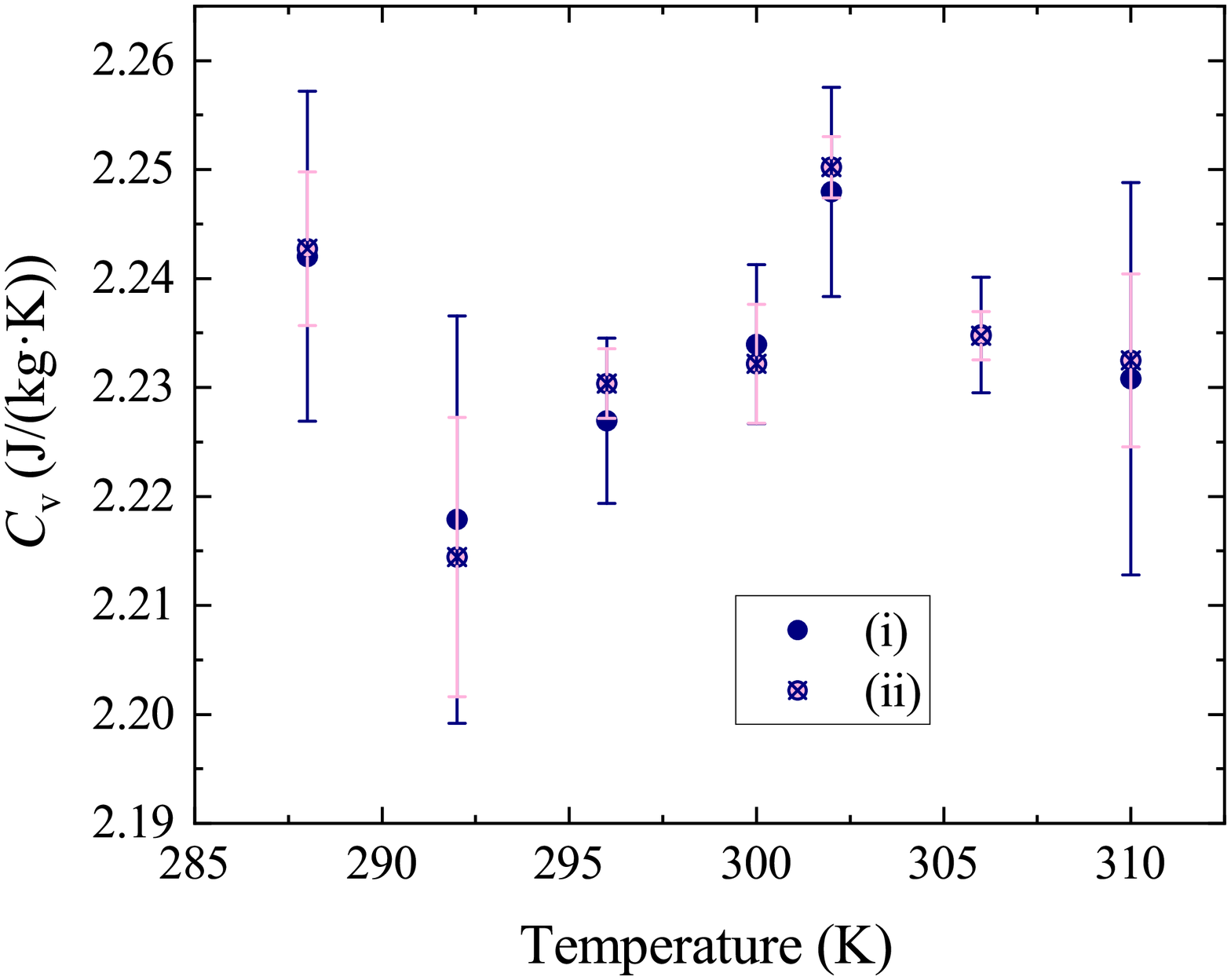}
\end{center}
\caption{\label{fig:heat_cap} Constant volume molar heat capacity $C_v$ as a function of temperature: i) total energy variation (Eq.~\ref{eq:heat_capacity_general}), and ii) total energy autocorrelation  (Eq.~\ref{eq:heat_capacity_time}).}
\end{figure}

On the other hand, for  ``thermostated systems'' one can obtain  constant volume heat capacity $C_v$ through the long-time averaging of $C_v(t)$ in steady-state regime~\cite{doi:10.1063/5.0046697}. The last one can be obtained in terms of energy correlation function, $C_E(t)$~\cite{Nielsen1996} with the use of fluctuation-dissipation theorem~\cite{2013i,1980iv,Ruscher2017}: 
\begin{equation}
 \label{eq:heat_capacity_time}
    C_v (t) = \frac{1}{k_BT^2}\left(C_E(0)-C_E(t)\right)
 \end{equation}
where $C_E(t)$ is the autocorrelation function of the total energy, defined as:
\begin{equation}
C_E(t) = \left< \delta E(t+t') \delta E(t')  \right>
\end{equation}
where $ \left<... \right>$ is the ensemble averaging (averaging over the $m$-independent configurations), $\delta E(t) = E(t) -  \left<E\right>$ is the total energy fluctuation term, and $<E>$ is the time- and ensemble-averaged value of the total energy. It is also important to underline that Eq.~\ref{eq:heat_capacity_time} is only relevant for long enough simulation times (higher than the typical relaxation time of the system~\cite{doi:10.1063/5.0046697}). Isochore heat capacities obtained by the two approaches (Eq.~\ref{eq:heat_capacity_general} and Eq.~\ref{eq:heat_capacity_time}) are plotted in Fig.~\ref{fig:heat_cap} for temperature in the range from $T \in [288,310]$ K. Both theoretical methods provide very similar results with relative discrepancies remain 10\%. Here, there is no noticeable variation of $C_v$ that might inform us about possible phase transition and reorganization of the atomic configuration. Besides, no experimental values are available to assess the reliability of the latter calculations. Thus, we will investigate constant pressure heat capacity $C_p$ for which differential calorimetry experiments were conducted.
Finally, taking into account well known equation from statistical physics that connects isobar heat capacity and enthalpy, $H$, one can obtain isobar heat capacity: 
\begin{equation}
 \label{eq:CP}
    C_p = \left(\frac{\partial H}{\partial T}\right)_p
 \end{equation}
where $H = E + pV$, where $p$ is the pressure of the system. 
To have a proper variation of $H$, we performed additional coolings in isobaric-isothermal ensemble for $t=2\times10^6$ fs. During this runs, each configuration at whole temperature set was cooled up to $T_f=270$ K with keeping the information about enthalpy and temperature change every $\delta t = 10^4$ fs. Then, a slope of $H$ vs. $T$ corresponds to the foreseen value of $C_p$.

\begin{figure}[!ht]
\begin{center}
\includegraphics[scale=0.5]{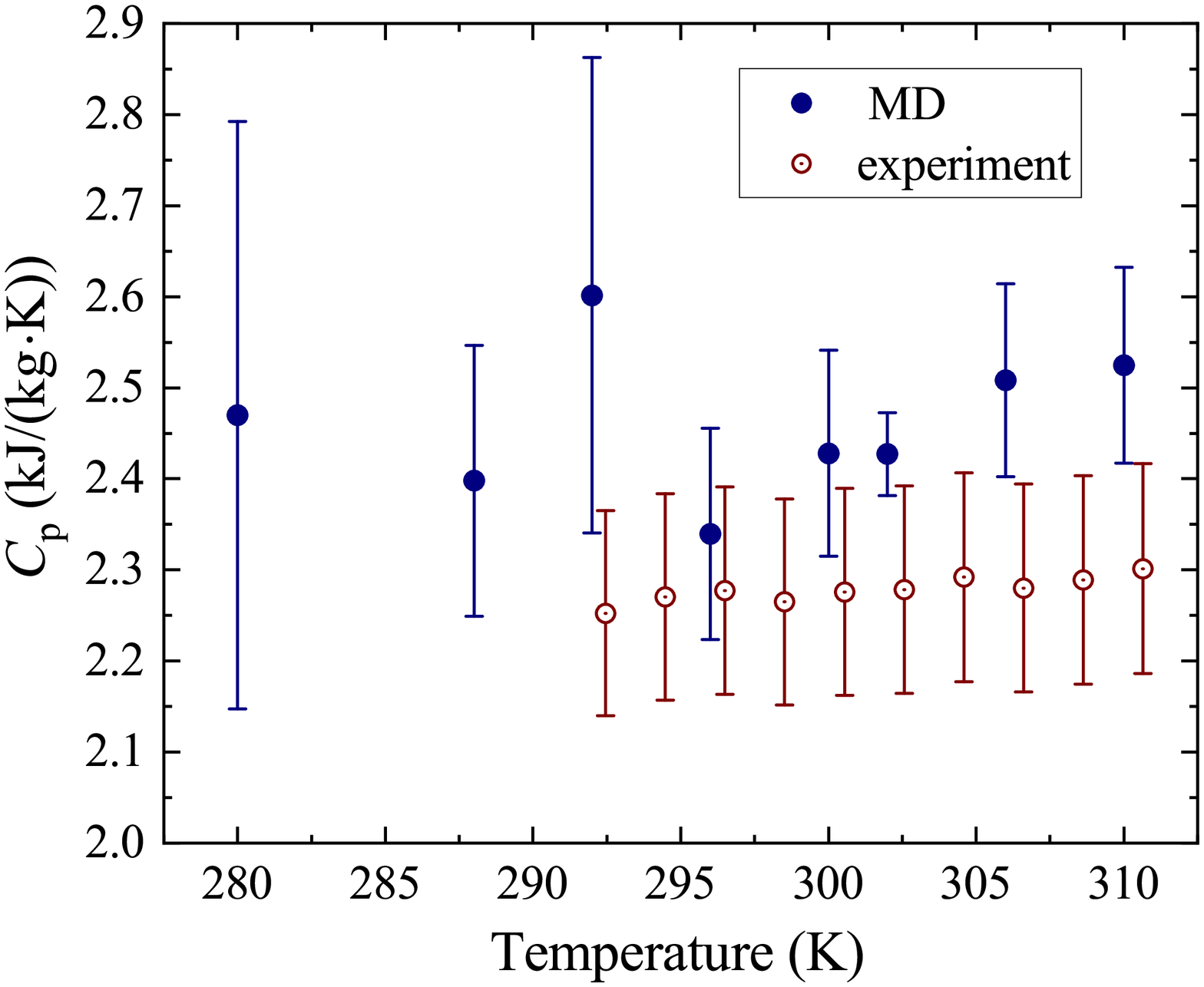}
\end{center}
\caption{\label{fig:CP} Temperature dependence of isobar molar heat capacity, $C_p(T)$,  of $n$-hexadecane molecules  obtained with MD simulations and experiments. }
\end{figure}
In Fig.~\ref{fig:CP} are reported constant pressure heat capacities experimentally and numerically. As previously, experiments were done in the liquid state while simulation data were evaluated from amorphous (280~K) to liquid (310~K) states. In addition to the overall good agreement between DSC and MD, one can clearly see a small peak in $C_p$ at 292~K, in a region where the liquid/amorphous transition occurs. It can be explained as follows: cooling toward samples state transition temperature leads to the breaking of ergodisity of the system.  That means that there is not enough time for the system to explore the phase space, and system's configurational degrees of freedom are not accessible anymore~\cite{CAVAGNA200951,Kauzmann1948}. In other words, the experimental time is smaller than the time required for an amorphous system to explore the phase space. The system is confined to the phase space's local energy minima with a decreased number of degrees of freedom compared to those that are available to the system at equilibrium and contribute to the specific heat. This explains why $C_p$ decreases dramatically while cooling the cooling temperature approaches state transition temperature $T_g$(and reaches approximately the same value in the crystal phase).

\subsection{Mass diffusivity}

The study of atom motions  can provide some useful insights about dynamical properties of the system. Among them, mass diffusivity can be related to the mean-square
displacement (MSD) of particles which can be obtained through time average of atom location as: 
\begin{equation}
\label{eq:MSD}
\text{MSD}(t) = \frac{1}{N}\sum_i\overline{\left( \boldsymbol{r}_i\left(t+t_0\right) - \boldsymbol{r}_i\left(t_0\right)\right)^{2}}
\end{equation}
where $\boldsymbol{r}_i\left(t_0\right)$ is the position of  $i^{\text{th}}$ particle, $N$ is the number of particles, $t_0$ is an initial time, and $\overline{...}$ corresponds to an average over the initial time $t_0$. For our system the calculation of MSD, over the center of mass of molecules, was performed. Under this assumption, $\boldsymbol{r}_i\left(t_0\right)$, and $N$ in Eq.~\ref{eq:MSD} were replaced by the position of center of mass of  $i^{\text{th}}$ molecule, and thus $N$ becomes the number of molecules.   
\begin{figure}[!ht]
\begin{center}
\includegraphics[scale=1]{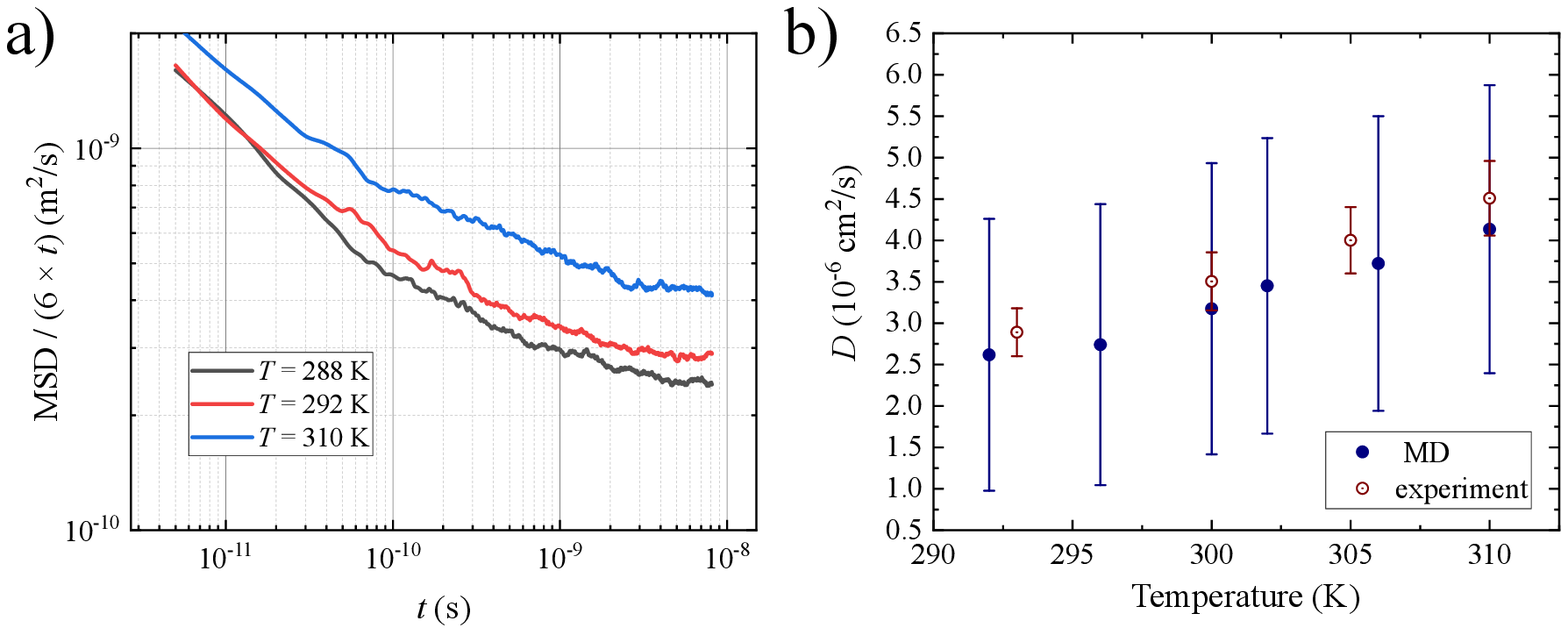}
\end{center}
\caption{\label{fig:msd} Temperature dependence of: a) MSD  and b) diffusion coefficient of center mass of $n$-hexadecane molecules obtained with MD simulations and NMR experiments.}
\end{figure}
 
When the system reaches steady state (i.e. for times $t$ longer compared to the relaxation time), MSD increases linearly with $t$ in a liquid as the diffusive motion dominates the ballistic one). This can be explained by slowing down of the dynamics with decreasing the temperature $T$, leading to emergence of an intermediate plateau regime~\cite{CAVAGNA200951}.  In this case the slope of MSD is proportional to the diffusion coefficient $D$:
\begin{equation}
\label{eq:D}
\text{MSD}(t) = 2dDt
\end{equation}
where $d=3$ is the space dimension of the considered system. Besides, in the case of MD simulations with periodic boundary conditions, it was found that the final value of diffusion coefficient, $D$, needs to be corrected by a term~\cite{Perego2020} which takes into account the size of the box and the shear viscosity. It reads: 
\begin{equation}
\label{eq:D_corrected}
 D = D_{\text{MSD}} + \frac{k_B T \xi}{6\pi\mu L}
\end{equation}
where $D_{\text{MSD}}$ is the diffusion coefficient of center of mass of $n$-hexadecane molecules that was calculated from the linear regime of MSD (i.e., slope of MSD$(t)$ in log-log scale is 1 for Eq.~\ref{eq:D}), $\xi \approx 2.837$ is a dimensionless constant~\cite{Yeh2004,Moultos2016,GuevaraCarrion2011}, $\mu$ is the dynamic viscosity, and $L$ is a box size. In our case the simulated system is a cube of length $L$, that depends on the chosen temperature: $L(T(K))$: $L(310)=6.315$ nm, $L(306)=6.306$ nm, $L(302)=6.295$ nm, $L(300)=6.291$ nm, $L(296)=6.281$ nm, and $L(292)=6.272$ nm.
Eq.~\ref{eq:D_corrected} is known as Yeh and Hummer (YH) relationship~\cite{Yeh2004,Moultos2016}. Experimental and numerically assessed value of hexadecane mass diffusivity $D$ as a function of temperature are shown in Fig.~\ref{fig:msd}. It can be noted that both - NMR experimental techniques and MD simulations agree even if MD results exhibit larger uncertainties. In addition, a clear increase of the mass diffusivity is observed while temperature increases as expected. 
 
\section{Conclusions}
\label{sec:conclusions}

This paper is devoted to the investigation of several quantities which describe thermal, mass and momentum transport in PCMs. More precisely, we studied thermophysical properties for $n$-hexadecane system at various temperatures using two frameworks - numerical and experimental one. The protocol developed for MD simulations and chosen force field allow us to evaluate these physical parameters for temperature range including liquid and amorphous states. It was presented that within numerical approach it is possible to determinate temperature region of liquid/amorphous transition.

Moreover, we demonstrate that the data for viscosity and other physical quantities such as isobar and isochor heat capacities, coefficient of thermal expansion, diffusion coefficient are all of the same order of magnitude between both methods and show good quantitative agreement between them. Calculated radial distribution function  is a good example of representing system ordering and packing at different $T$'s.  

From experimental point of view, with the use of nuclear magnetic resonance (NMR) we obtained diffusion coefficient at different temperatures. The calculated results are in a good agreement with ones computed by molecular dynamics simulations. Furthermore, temperature dependence of constant pressure heat capacity in a liquid regime shows a good agreement between   diferential  scanning  calorimetry (DSC) and MD. 

With the numerical framework it was observed liquid-amorphous state transition for temperature $T_g \in \left(296,302\right)$ K. Such transition can be seen from occurring peaks for $\kappa$ or $C_p$. Moreover, the drastic increase of viscosity with increasing temperature can serve as one more argument for system experiences a transition. 

As a conclusion, it is important to note that all the simulations were time consuming. Moreover, MD simulations accurately depict the behavior of the liquid phase near the phase transition point of PCMs, and the obtained results can serve as a foundation for further research of the PCMs-based nanocomposite characteristics. Further investigations will have to adress PCM crystallization. This remains a computing challenge which we intend to tackle in future works. 

\section{Acknwolegement}
This paper contains the results obtained in the frames of the project ``Hotline'' ANR-19-CE09-0003 and DropSurf ANR-20-CE05-0030. This work was performed using HPC resources from GENCI-TGCC and GENCI-IDRIS (2021-A0110913052), in addition HPC resources were partially provided by the EXPLOR centre hosted by the Universitt\'e de Lorraine.  Thanks to “STOCK NRJ” that is co-financed by the European Union within the framework of the Program FEDER-FSE Lorraine and Massif des Vosges 2014–2020.

\clearpage
\newpage

\bibliographystyle{unsrt}

\bibliography{references.bib}

\providecommand{\noopsort}[1]{}\providecommand{\singleletter}[1]{#1}%
\begin{thebibliography}{10}
\expandafter\ifx\csname url\endcsname\relax
  \def\url#1{\texttt{#1}}\fi
\expandafter\ifx\csname urlprefix\endcsname\relax\def\urlprefix{URL }\fi
\expandafter\ifx\csname href\endcsname\relax
  \def\href#1#2{#2} \def\path#1{#1}\fi

\bibitem{Hu2008}
J.~Hu, K.~Babu, \href{https://doi.org/10.1533/9781845695064.275}{Testing
  intelligent textiles}, in: Fabric Testing, Elsevier, 2008, pp. 275--308.
\newblock \href {http://dx.doi.org/10.1533/9781845695064.275}
  {\path{doi:10.1533/9781845695064.275}}.
\newline\urlprefix\url{https://doi.org/10.1533/9781845695064.275}

\bibitem{Ho2011}
C.~Ho, J.~Fan, E.~Newton, R.~Au,
  \href{https://doi.org/10.1533/9780857090645.2.165}{Improving thermal comfort
  in apparel}, in: Improving Comfort in Clothing, Elsevier, 2011, pp. 165--181.
\newblock \href {http://dx.doi.org/10.1533/9780857090645.2.165}
  {\path{doi:10.1533/9780857090645.2.165}}.
\newline\urlprefix\url{https://doi.org/10.1533/9780857090645.2.165}

\bibitem{Chaturvedi2021}
R.~Chaturvedi, A.~Islam, K.~Sharma,
  \href{https://doi.org/10.1016/j.matpr.2020.11.665}{A review on the
  applications of {PCM} in thermal storage of solar energy}, Materials Today:
  Proceedings 43 (2021) 293--297.
\newblock \href {http://dx.doi.org/10.1016/j.matpr.2020.11.665}
  {\path{doi:10.1016/j.matpr.2020.11.665}}.
\newline\urlprefix\url{https://doi.org/10.1016/j.matpr.2020.11.665}

\bibitem{AbbasiKamazani2021}
M.~A. Kamazani, C.~Aghanajafi, \href{https://doi.org/10.1002/er.7279}{Numerical
  simulation of geothermal-{PVT} hybrid system with {PCM} storage tank},
  International Journal of Energy Research\href
  {http://dx.doi.org/10.1002/er.7279} {\path{doi:10.1002/er.7279}}.
\newline\urlprefix\url{https://doi.org/10.1002/er.7279}

\bibitem{Yao2020}
J.~Yao, P.~Zhu, L.~Guo, L.~Duan, Z.~Zhang, S.~Kurko, Z.~Wu,
  \href{https://doi.org/10.1016/j.ijhydene.2020.05.089}{A continuous hydrogen
  absorption/desorption model for metal hydride reactor coupled with {PCM} as
  heat management and its application in the fuel cell power system},
  International Journal of Hydrogen Energy 45~(52) (2020) 28087--28099.
\newblock \href {http://dx.doi.org/10.1016/j.ijhydene.2020.05.089}
  {\path{doi:10.1016/j.ijhydene.2020.05.089}}.
\newline\urlprefix\url{https://doi.org/10.1016/j.ijhydene.2020.05.089}

\bibitem{Cheng2021}
P.~Cheng, X.~Chen, H.~Gao, X.~Zhang, Z.~Tang, A.~Li, G.~Wang,
  \href{https://doi.org/10.1016/j.nanoen.2021.105948}{Different dimensional
  nanoadditives for thermal conductivity enhancement of phase change materials:
  Fundamentals and applications}, Nano Energy 85 (2021) 105948.
\newblock \href {http://dx.doi.org/10.1016/j.nanoen.2021.105948}
  {\path{doi:10.1016/j.nanoen.2021.105948}}.
\newline\urlprefix\url{https://doi.org/10.1016/j.nanoen.2021.105948}

\bibitem{Aljaerani2022}
H.~A. Aljaerani, M.~Samykano, A.~K. Pandey, K.~Kadirgama, M.~George, R.~Saidur,
  \href{https://doi.org/10.1016/j.icheatmasstransfer.2022.105898}{{Thermophysical
  properties enhancement and characterization of CuO nanoparticles enhanced
  HITEC molten salt for concentrated solar power applications}}, International
  Communications in Heat and Mass Transfer 132~(January) (2022) 105898.
\newblock \href {http://dx.doi.org/10.1016/j.icheatmasstransfer.2022.105898}
  {\path{doi:10.1016/j.icheatmasstransfer.2022.105898}}.
\newline\urlprefix\url{https://doi.org/10.1016/j.icheatmasstransfer.2022.105898}

\bibitem{Esapour2018}
M.~Esapour, A.~Hamzehnezhad, A.~A.~R. Darzi, M.~Jourabian,
  \href{https://doi.org/10.1016/j.enconman.2018.05.086}{Melting and
  solidification of {PCM} embedded in porous metal foam in horizontal
  multi-tube heat storage system}, Energy Conversion and Management 171 (2018)
  398--410.
\newblock \href {http://dx.doi.org/10.1016/j.enconman.2018.05.086}
  {\path{doi:10.1016/j.enconman.2018.05.086}}.
\newline\urlprefix\url{https://doi.org/10.1016/j.enconman.2018.05.086}

\bibitem{Qureshi2021}
Z.~A. Qureshi, S.~A.~B. Al-Omari, E.~Elnajjar, O.~Al-Ketan, R.~A. Al-Rub,
  \href{https://doi.org/10.1016/j.icheatmasstransfer.2021.105265}{{Using triply
  periodic minimal surfaces (TPMS)-based metal foams structures as skeleton for
  metal-foam-PCM composites for thermal energy storage and energy management
  applications}}, International Communications in Heat and Mass Transfer 124
  (2021) 105265.
\newblock \href {http://dx.doi.org/10.1016/j.icheatmasstransfer.2021.105265}
  {\path{doi:10.1016/j.icheatmasstransfer.2021.105265}}.
\newline\urlprefix\url{https://doi.org/10.1016/j.icheatmasstransfer.2021.105265}

\bibitem{Qiu2021}
L.~Qiu, K.~Yan, Y.~Feng, X.~Liu, X.~Zhang,
  \href{https://doi.org/10.1016/j.coco.2021.100892}{Bionic hierarchical porous
  aluminum nitride ceramic composite phase change material with excellent heat
  transfer and storage performance}, Composites Communications 27 (2021)
  100892.
\newblock \href {http://dx.doi.org/10.1016/j.coco.2021.100892}
  {\path{doi:10.1016/j.coco.2021.100892}}.
\newline\urlprefix\url{https://doi.org/10.1016/j.coco.2021.100892}

\bibitem{Nakhchi2020}
M.~Nakhchi, J.~Esfahani,
  \href{https://doi.org/10.1016/j.est.2020.101424}{Improving the melting
  performance of {PCM} thermal energy storage with novel stepped fins}, Journal
  of Energy Storage 30 (2020) 101424.
\newblock \href {http://dx.doi.org/10.1016/j.est.2020.101424}
  {\path{doi:10.1016/j.est.2020.101424}}.
\newline\urlprefix\url{https://doi.org/10.1016/j.est.2020.101424}

\bibitem{Safari2017}
A.~Safari, R.~Saidur, F.~Sulaiman, Y.~Xu, J.~Dong,
  \href{https://doi.org/10.1016/j.rser.2016.11.272}{A review on supercooling of
  phase change materials in thermal energy storage systems}, Renewable and
  Sustainable Energy Reviews 70 (2017) 905--919.
\newblock \href {http://dx.doi.org/10.1016/j.rser.2016.11.272}
  {\path{doi:10.1016/j.rser.2016.11.272}}.
\newline\urlprefix\url{https://doi.org/10.1016/j.rser.2016.11.272}

\bibitem{Beaupere2018}
N.~Beaupere, U.~Soupremanien, L.~Zalewski,
  \href{https://doi.org/10.1016/j.tca.2018.10.009}{Nucleation triggering
  methods in supercooled phase change materials ({PCM}), a review},
  Thermochimica Acta 670 (2018) 184--201.
\newblock \href {http://dx.doi.org/10.1016/j.tca.2018.10.009}
  {\path{doi:10.1016/j.tca.2018.10.009}}.
\newline\urlprefix\url{https://doi.org/10.1016/j.tca.2018.10.009}

\bibitem{Isaiev2020}
M.~Isaiev, X.~Wang, K.~Termentzidis, D.~Lacroix,
  \href{https://doi.org/10.1063/5.0014680}{Thermal transport enhancement of
  hybrid nanocomposites; impact of confined water inside nanoporous silicon},
  Applied Physics Letters 117~(3) (2020) 033701.
\newblock \href {http://dx.doi.org/10.1063/5.0014680}
  {\path{doi:10.1063/5.0014680}}.
\newline\urlprefix\url{https://doi.org/10.1063/5.0014680}

\bibitem{Zhao2020}
C.~Zhao, Y.~Tao, Y.~Yu,
  \href{https://doi.org/10.1016/j.ijheatmasstransfer.2020.119382}{Molecular
  dynamics simulation of nanoparticle effect on melting enthalpy of paraffin
  phase change material}, International Journal of Heat and Mass Transfer 150
  (2020) 119382.
\newblock \href {http://dx.doi.org/10.1016/j.ijheatmasstransfer.2020.119382}
  {\path{doi:10.1016/j.ijheatmasstransfer.2020.119382}}.
\newline\urlprefix\url{https://doi.org/10.1016/j.ijheatmasstransfer.2020.119382}

\bibitem{Yan2021}
X.~Yan, Y.~Feng, L.~Qiu, X.~Zhang,
  \href{https://doi.org/10.1016/j.energy.2021.121158}{Thermal conductivity and
  phase change characteristics of hierarchical porous diamond/erythritol
  composite phase change materials}, Energy 233 (2021) 121158.
\newblock \href {http://dx.doi.org/10.1016/j.energy.2021.121158}
  {\path{doi:10.1016/j.energy.2021.121158}}.
\newline\urlprefix\url{https://doi.org/10.1016/j.energy.2021.121158}

\bibitem{LuningPrak2021}
D.~J.~L. Prak, J.~A. Harrison, B.~H. Morrow,
  \href{https://doi.org/10.1021/acs.jced.0c01043}{Thermophysical properties of
  two-component mixtures of n-nonylbenzene or 1, 3, 5-triisopropylbenzene with
  n-hexadecane or n-dodecane at 0.1 {MPa}: Experimentally measured densities,
  viscosities, and speeds of sound and molecular packing modeled using
  molecular dynamics simulations}, Journal of Chemical {\&} Engineering Data
  66~(3) (2021) 1442--1456.
\newblock \href {http://dx.doi.org/10.1021/acs.jced.0c01043}
  {\path{doi:10.1021/acs.jced.0c01043}}.
\newline\urlprefix\url{https://doi.org/10.1021/acs.jced.0c01043}

\bibitem{Zhang2020}
M.~Zhang, C.~Wang, A.~Luo, Z.~Liu, X.~Zhang,
  \href{https://doi.org/10.1016/j.applthermaleng.2019.114639}{Molecular
  dynamics simulation on thermophysics of paraffin/{EVA}/graphene
  nanocomposites as phase change materials}, Applied Thermal Engineering 166
  (2020) 114639.
\newblock \href {http://dx.doi.org/10.1016/j.applthermaleng.2019.114639}
  {\path{doi:10.1016/j.applthermaleng.2019.114639}}.
\newline\urlprefix\url{https://doi.org/10.1016/j.applthermaleng.2019.114639}

\bibitem{Zhao2022}
C.~Y. Zhao, Y.~B. Tao, Y.~S. Yu, {Thermal conductivity enhancement of phase
  change material with charged nanoparticle: A molecular dynamics simulation},
  Energy 242 (2022) 1--8.
\newblock \href {http://dx.doi.org/10.1016/j.energy.2021.123033}
  {\path{doi:10.1016/j.energy.2021.123033}}.

\bibitem{Morrow2021}
J.~A. Morrow, Brian H.and~Harrison,
  \href{https://doi.org/10.1021/acs.energyfuels.0c03363}{Evaluating the ability
  of selected force fields to simulate hydrocarbons as a function of
  temperature and pressure using molecular dynamics}, Energy \& Fuels 35~(5)
  (2021) 3742--3752.
\newblock \href {http://dx.doi.org/10.1021/acs.energyfuels.0c03363}
  {\path{doi:10.1021/acs.energyfuels.0c03363}}.
\newline\urlprefix\url{https://doi.org/10.1021/acs.energyfuels.0c03363}

\bibitem{Zhao2021}
C.~Zhao, Y.~Tao, Y.~Yu,
  \href{https://doi.org/10.1016/j.molliq.2021.115448}{Molecular dynamics
  simulation of thermal and phonon transport characteristics of nanocomposite
  phase change material}, Journal of Molecular Liquids 329 (2021) 115448.
\newblock \href {http://dx.doi.org/10.1016/j.molliq.2021.115448}
  {\path{doi:10.1016/j.molliq.2021.115448}}.
\newline\urlprefix\url{https://doi.org/10.1016/j.molliq.2021.115448}

\bibitem{Morrow2018}
B.~H. Morrow, S.~Maskey, M.~Z. Gustafson, D.~J.~L. Prak, J.~A. Harrison,
  \href{https://doi.org/10.1021/acs.jpcb.8b03752}{Impact of molecular structure
  on properties of n-hexadecane and alkylbenzene binary mixtures}, The Journal
  of Physical Chemistry B 122~(25) (2018) 6595--6603.
\newblock \href {http://dx.doi.org/10.1021/acs.jpcb.8b03752}
  {\path{doi:10.1021/acs.jpcb.8b03752}}.
\newline\urlprefix\url{https://doi.org/10.1021/acs.jpcb.8b03752}

\bibitem{cabeza2015unconventional}
L.~F. Cabeza, C.~Barreneche, I.~Martorell, L.~Mir{\'o}, S.~Sari-Bey, M.~Fois,
  H.~O. Paksoy, N.~Sahan, R.~Weber, M.~Constantinescu, et~al., Unconventional
  experimental technologies available for phase change materials (pcm)
  characterization. part 1. thermophysical properties, Renewable and
  Sustainable Energy Reviews 43 (2015) 1399--1414.

\bibitem{ma12182974}
M.~Faden, S.~Höhlein, J.~Wanner, A.~König-Haagen, D.~Brüggemann,
  \href{https://www.mdpi.com/1996-1944/12/18/2974}{Review of thermophysical
  property data of octadecane for phase-change studies}, Materials 12~(18).
\newblock \href {http://dx.doi.org/10.3390/ma12182974}
  {\path{doi:10.3390/ma12182974}}.
\newline\urlprefix\url{https://www.mdpi.com/1996-1944/12/18/2974}

\bibitem{fernandez2015unconventional}
A.~I. Fern{\'a}ndez, A.~Sol{\'e}, J.~Gir{\'o}-Paloma, M.~Mart{\'\i}nez,
  M.~Hadjieva, A.~Boudenne, M.~Constantinescu, E.~M. Anghel, M.~Malikova,
  I.~Krupa, et~al., Unconventional experimental technologies used for phase
  change materials (pcm) characterization: part 2--morphological and structural
  characterization, physico-chemical stability and mechanical properties,
  Renewable and Sustainable Energy Reviews 43 (2015) 1415--1426.

\bibitem{Sgreva2022}
N.~R. Sgreva, J.~Noel, C.~M{\'{e}}tivier, P.~Marchal, H.~Chaynes, M.~Isaiev,
  Y.~Jannot, \href{https://doi.org/10.1016/j.tca.2022.179180}{Thermo-physical
  characterization of hexadecane during the solid/liquid phase change},
  Thermochimica Acta\href {http://dx.doi.org/10.1016/j.tca.2022.179180}
  {\path{doi:10.1016/j.tca.2022.179180}}.
\newline\urlprefix\url{https://doi.org/10.1016/j.tca.2022.179180}

\bibitem{Noel2022}
J.~Noel, Y.~Jannot, C.~Métivier, N.~Sgreva, Thermal properties of polyethylene
  glycol 600 in liquid, solid phases and through the phase transitionn, under
  revisions for thermochimica acta.

\bibitem{Mitsuhashi2010}
R.~Mitsuhashi, Y.~Suzuki, Y.~Yamanari, H.~Mitamura, T.~Kambe, N.~Ikeda,
  H.~Okamoto, A.~Fujiwara, M.~Yamaji, N.~Kawasaki, Y.~Maniwa, Y.~Kubozono,
  \href{https://doi.org/10.1038/nature08859}{Superconductivity in
  alkali-metal-doped picene}, Nature 464~(7285) (2010) 76--79.
\newblock \href {http://dx.doi.org/10.1038/nature08859}
  {\path{doi:10.1038/nature08859}}.
\newline\urlprefix\url{https://doi.org/10.1038/nature08859}

\bibitem{doi:10.1126/science.1148326}
A.~Tuteja, W.~Choi, M.~Ma, J.~M. Mabry, S.~A. Mazzella, G.~C. Rutledge, G.~H.
  McKinley, R.~E. Cohen, Designing superoleophobic surfaces, Science 318~(5856)
  (2007) 1618--1622.
\newblock \href {http://dx.doi.org/10.1126/science.1148326}
  {\path{doi:10.1126/science.1148326}}.

\bibitem{Chriaa2020}
I.~Chriaa, A.~Trigui, M.~Karkri, I.~Jedidi, M.~Abdelmouleh, C.~Boudaya,
  \href{https://doi.org/10.1016/j.applthermaleng.2020.115072}{Thermal
  properties of shape-stabilized phase change materials based on low density
  polyethylene, hexadecane and {SEBS} for thermal energy storage}, Applied
  Thermal Engineering 171 (2020) 115072.
\newblock \href {http://dx.doi.org/10.1016/j.applthermaleng.2020.115072}
  {\path{doi:10.1016/j.applthermaleng.2020.115072}}.
\newline\urlprefix\url{https://doi.org/10.1016/j.applthermaleng.2020.115072}

\bibitem{tuckerman2010statistical}
M.~Tuckerman, \href{https://books.google.fr/books?id=Lo3Jqc0pgrcC}{Statistical
  Mechanics: Theory and Molecular Simulation}, Oxford Graduate Texts, OUP
  Oxford, 2010.
\newline\urlprefix\url{https://books.google.fr/books?id=Lo3Jqc0pgrcC}

\bibitem{Lenahan2021}
F.~D. Lenahan, M.~Zikeli, M.~H. Rausch, T.~Klein, A.~P. Fr{\"{o}}ba,
  {Viscosity, Interfacial Tension, and Density of Binary-Liquid Mixtures of
  n-Hexadecane with n-Octacosane, 2,2,4,4,6,8,8-Heptamethylnonane, or
  1-Hexadecanol at Temperatures between 298.15 and 573.15 K by Surface Light
  Scattering and Equilibrium Molecular Dynamics Simulations}, Journal of
  Chemical and Engineering Data 66~(5) (2021) 2264--2280.
\newblock \href {http://dx.doi.org/10.1021/acs.jced.1c00108}
  {\path{doi:10.1021/acs.jced.1c00108}}.

\bibitem{DaSilva2021}
G.~C. da~Silva, F.~G. Oliveira, W.~F. de~Souza, M.~C. de~Oliveira, P.~M.
  Esteves, B.~A. Horta,
  \href{https://doi.org/10.1016/j.fuel.2020.119029}{{Effects of paraffin, fatty
  acid and long alkyl chain phenol on the solidification of n-hexadecane under
  harsh subcooling condition: A molecular dynamics simulation study}}, Fuel
  285~(September 2020) (2021) 119029.
\newblock \href {http://dx.doi.org/10.1016/j.fuel.2020.119029}
  {\path{doi:10.1016/j.fuel.2020.119029}}.
\newline\urlprefix\url{https://doi.org/10.1016/j.fuel.2020.119029}

\bibitem{Sharma2022}
A.~Sharma, P.~Parth, S.~Shobhana, M.~Bobin, B.~K. Hardik,
  \href{https://doi.org/10.1016/j.icheatmasstransfer.2021.105792}{{Numerical
  study of ice freezing process on fin aided thermal energy storage system}},
  International Communications in Heat and Mass Transfer 130 (2022) 105792.
\newblock \href {http://dx.doi.org/10.1016/j.icheatmasstransfer.2021.105792}
  {\path{doi:10.1016/j.icheatmasstransfer.2021.105792}}.
\newline\urlprefix\url{https://doi.org/10.1016/j.icheatmasstransfer.2021.105792}

\bibitem{Surblys2019}
D.~Surblys, H.~Matsubara, G.~Kikugawa, T.~Ohara,
  \href{https://doi.org/10.1103/physreve.99.051301}{Application of atomic
  stress to compute heat flux via molecular dynamics for systems with many-body
  interactions}, Physical Review E 99~(5).
\newblock \href {http://dx.doi.org/10.1103/physreve.99.051301}
  {\path{doi:10.1103/physreve.99.051301}}.
\newline\urlprefix\url{https://doi.org/10.1103/physreve.99.051301}

\bibitem{Surblys2021}
D.~Surblys, H.~Matsubara, G.~Kikugawa, T.~Ohara, {Methodology and meaning of
  computing heat flux via atomic stress in systems with constraint dynamics},
  Journal of Applied Physics 130~(21).
\newblock \href {http://dx.doi.org/10.1063/5.0070930}
  {\path{doi:10.1063/5.0070930}}.

\bibitem{Espeau1996}
P.~Espeau, L.~Robles, D.~Mondieig, Y.~Haget, M.~Cuevas-Diarte, H.~Oonk, Mise au
  point sur le comportement {\'e}nerg{\'e}tique et cristallographique des
  n-alcanes-i. s{\'e}rie de c8h18 {\`a} c21h44, Journal de chimie physique 93
  (1996) 1217--1238.

\bibitem{VELEZ2015383}
C.~Vélez, M.~Khayet, J.~{Ortiz de Zárate},
  \href{https://www.sciencedirect.com/science/article/pii/S0306261915000707}{Temperature-dependent
  thermal properties of solid/liquid phase change even-numbered n-alkanes:
  n-hexadecane, n-octadecane and n-eicosane}, Applied Energy 143 (2015)
  383--394.
\newblock \href
  {http://dx.doi.org/https://doi.org/10.1016/j.apenergy.2015.01.054}
  {\path{doi:https://doi.org/10.1016/j.apenergy.2015.01.054}}.
\newline\urlprefix\url{https://www.sciencedirect.com/science/article/pii/S0306261915000707}

\bibitem{jannot_thermal_2018}
Y.~Jannot, A.~Degiovanni,
  \href{https://hal.univ-lorraine.fr/hal-01684518}{Thermal properties
  measurement of materials}, ISTE ; WILEY, 2018, pages: 342 pages.
\newline\urlprefix\url{https://hal.univ-lorraine.fr/hal-01684518}

\bibitem{Tanner1970}
J.~E. Tanner, \href{https://doi.org/10.1063/1.1673336}{Use of the stimulated
  echo in {NMR} diffusion studies}, The Journal of Chemical Physics 52~(5)
  (1970) 2523--2526.
\newblock \href {http://dx.doi.org/10.1063/1.1673336}
  {\path{doi:10.1063/1.1673336}}.
\newline\urlprefix\url{https://doi.org/10.1063/1.1673336}

\bibitem{stejskal_spin_1965}
E.~O. Stejskal, J.~E. Tanner,
  \href{http://aip.scitation.org/doi/10.1063/1.1695690}{Spin {Diffusion}
  {Measurements}: {Spin} {Echoes} in the {Presence} of a {Time}‐{Dependent}
  {Field} {Gradient}}, The Journal of Chemical Physics 42~(1) (1965) 288--292.
\newblock \href {http://dx.doi.org/10.1063/1.1695690}
  {\path{doi:10.1063/1.1695690}}.
\newline\urlprefix\url{http://aip.scitation.org/doi/10.1063/1.1695690}

\bibitem{PACKMOL}
L.~Martínez, R.~Andrade, E.~G. Birgin, J.~M. Martínez,
  \href{https://onlinelibrary.wiley.com/doi/abs/10.1002/jcc.21224}{Packmol: A
  package for building initial configurations for molecular dynamics
  simulations}, Journal of Computational Chemistry 30~(13) (2009) 2157--2164.
\newblock \href
  {http://arxiv.org/abs/https://onlinelibrary.wiley.com/doi/pdf/10.1002/jcc.21224}
  {\path{arXiv:https://onlinelibrary.wiley.com/doi/pdf/10.1002/jcc.21224}},
  \href {http://dx.doi.org/https://doi.org/10.1002/jcc.21224}
  {\path{doi:https://doi.org/10.1002/jcc.21224}}.
\newline\urlprefix\url{https://onlinelibrary.wiley.com/doi/abs/10.1002/jcc.21224}

\bibitem{HUMP96}
W.~Humphrey, A.~Dalke, K.~Schulten, {VMD} -- {V}isual {M}olecular {D}ynamics,
  Journal of Molecular Graphics 14 (1996) 33--38.

\bibitem{STON1998}
J.~Stone, {\em An Efficient Library for Parallel Ray Tracing and Animation},
  Master's thesis, Computer Science Department, University of Missouri-Rolla
  (April 1998).

\bibitem{LAMMPS}
Large-scale atomic/molecular massively parallel simulator,
  \url{https://www.lammps.org}.

\bibitem{LOPLS}
S.~W.~I. Siu, K.~Pluhackova, R.~A. Böckmann,
  \href{https://doi.org/10.1021/ct200908r}{Optimization of the opls-aa force
  field for long hydrocarbons}, Journal of Chemical Theory and Computation
  8~(4) (2012) 1459--1470, pMID: 26596756.
\newblock \href {http://arxiv.org/abs/https://doi.org/10.1021/ct200908r}
  {\path{arXiv:https://doi.org/10.1021/ct200908r}}, \href
  {http://dx.doi.org/10.1021/ct200908r} {\path{doi:10.1021/ct200908r}}.
\newline\urlprefix\url{https://doi.org/10.1021/ct200908r}

\bibitem{Moultos2016}
O.~A. Moultos, Y.~Zhang, I.~N. Tsimpanogiannis, I.~G. Economou, E.~J. Maginn,
  \href{https://doi.org/10.1063/1.4960776}{System-size corrections for
  self-diffusion coefficients calculated from molecular dynamics simulations:
  The case of {CO}2, n-alkanes, and poly(ethylene glycol) dimethyl ethers}, The
  Journal of Chemical Physics 145~(7) (2016) 074109.
\newblock \href {http://dx.doi.org/10.1063/1.4960776}
  {\path{doi:10.1063/1.4960776}}.
\newline\urlprefix\url{https://doi.org/10.1063/1.4960776}

\bibitem{Papavasileiou2019}
K.~D. Papavasileiou, L.~D. Peristeras, A.~Bick, I.~G. Economou,
  \href{https://doi.org/10.1021/acs.jpcb.9b02840}{Molecular dynamics simulation
  of pure n-alkanes and their mixtures at elevated temperatures using atomistic
  and coarse-grained force fields}, The Journal of Physical Chemistry B
  123~(29) (2019) 6229--6243.
\newblock \href {http://dx.doi.org/10.1021/acs.jpcb.9b02840}
  {\path{doi:10.1021/acs.jpcb.9b02840}}.
\newline\urlprefix\url{https://doi.org/10.1021/acs.jpcb.9b02840}

\bibitem{Guido2019}
A.~David, A.~De~Nicola, U.~Tartaglino, G.~Milano, G.~Raos, Viscoelasticity of
  short polymer liquids from atomistic simulations, Journal of The
  Electrochemical Society 166 (2019) B3246--B3256.
\newblock \href {http://dx.doi.org/10.1149/2.0371909jes}
  {\path{doi:10.1149/2.0371909jes}}.

\bibitem{Siu2012}
S.~W.~I. Siu, K.~Pluhackova, R.~A. B{\"o}ckmann,
  \href{https://doi.org/10.1021/ct200908r}{Optimization of the opls-aa force
  field for long hydrocarbons}, Journal of Chemical Theory and Computation
  8~(4) (2012) 1459--1470.
\newblock \href {http://dx.doi.org/10.1021/ct200908r}
  {\path{doi:10.1021/ct200908r}}.
\newline\urlprefix\url{https://doi.org/10.1021/ct200908r}

\bibitem{RYCKAERT1977327}
J.-P. Ryckaert, G.~Ciccotti, H.~J. Berendsen, Numerical integration of the
  cartesian equations of motion of a system with constraints: molecular
  dynamics of n-alkanes, Journal of Computational Physics 23~(3) (1977)
  327--341.

\bibitem{Nose:JCP1984}
S.~Nosé, \href{https://doi.org/10.1063/1.447334}{A unified formulation of the
  constant temperature molecular dynamics methods}, The Journal of Chemical
  Physics 81~(1) (1984) 511--519.
\newblock \href {http://arxiv.org/abs/https://doi.org/10.1063/1.447334}
  {\path{arXiv:https://doi.org/10.1063/1.447334}}, \href
  {http://dx.doi.org/10.1063/1.447334} {\path{doi:10.1063/1.447334}}.
\newline\urlprefix\url{https://doi.org/10.1063/1.447334}

\bibitem{Lal2000}
K.~Lal, N.~Tripathi, G.~P. Dubey,
  \href{https://doi.org/10.1021/je000103x}{Densities, viscosities, and
  refractive indices of binary liquid mixtures of hexane, decane, hexadecane,
  and squalane with benzene at 298.15 k}, Journal of Chemical {\&} Engineering
  Data 45~(5) (2000) 961--964.
\newblock \href {http://dx.doi.org/10.1021/je000103x}
  {\path{doi:10.1021/je000103x}}.
\newline\urlprefix\url{https://doi.org/10.1021/je000103x}

\bibitem{doi:10.1021/acs.jctc.9b00252}
P.~Boone, H.~Babaei, C.~E. Wilmer, Heat flux for many-body interactions:
  Corrections to lammps, Journal of Chemical Theory and Computation 15~(10)
  (2019) 5579--5587, pMID: 31369260.
\newblock \href {http://dx.doi.org/10.1021/acs.jctc.9b00252}
  {\path{doi:10.1021/acs.jctc.9b00252}}.

\bibitem{PhysRevE.99.051301}
D.~Surblys, H.~Matsubara, G.~Kikugawa, T.~Ohara,
  \href{https://link.aps.org/doi/10.1103/PhysRevE.99.051301}{Application of
  atomic stress to compute heat flux via molecular dynamics for systems with
  many-body interactions}, Phys. Rev. E 99 (2019) 051301.
\newblock \href {http://dx.doi.org/10.1103/PhysRevE.99.051301}
  {\path{doi:10.1103/PhysRevE.99.051301}}.
\newline\urlprefix\url{https://link.aps.org/doi/10.1103/PhysRevE.99.051301}

\bibitem{Isaeva2019}
L.~Isaeva, G.~Barbalinardo, D.~Donadio, S.~Baroni,
  \href{http://www.nature.com/articles/s41467-019-11572-4}{{Modeling heat
  transport in crystals and glasses from a unified lattice-dynamical
  approach}}, Nature Communications 10~(1) (2019) 3853.
\newblock \href {http://dx.doi.org/10.1038/s41467-019-11572-4}
  {\path{doi:10.1038/s41467-019-11572-4}}.
\newline\urlprefix\url{http://www.nature.com/articles/s41467-019-11572-4}

\bibitem{DosSantos2013}
W.~N. {Dos Santos}, J.~A. {De Sousa}, R.~Gregorio,
  \href{http://dx.doi.org/10.1016/j.polymertesting.2013.05.007}{{Thermal
  conductivity behaviour of polymers around glass transition and crystalline
  melting temperatures}}, Polymer Testing 32~(5) (2013) 987--994.
\newblock \href {http://dx.doi.org/10.1016/j.polymertesting.2013.05.007}
  {\path{doi:10.1016/j.polymertesting.2013.05.007}}.
\newline\urlprefix\url{http://dx.doi.org/10.1016/j.polymertesting.2013.05.007}

\bibitem{inproceedings}
D.~Vasco, M.~Mu\~{n}oz, P.~Galvez, P.~A. Zapata, Modification of cu and cuo
  nanoparticles with oleic acid and thermal characterization of paraffins:
  towards the preparation of a stable nanopcm, 2015.

\bibitem{doi:10.1063/5.0046697}
L.~Klochko, J.~Baschnagel, J.~P. Wittmer, A.~N. Semenov,
  \href{https://doi.org/10.1063/5.0046697}{General relations to obtain the
  time-dependent heat capacity from isothermal simulations}, The Journal of
  Chemical Physics 154~(16) (2021) 164501.
\newblock \href {http://arxiv.org/abs/https://doi.org/10.1063/5.0046697}
  {\path{arXiv:https://doi.org/10.1063/5.0046697}}, \href
  {http://dx.doi.org/10.1063/5.0046697} {\path{doi:10.1063/5.0046697}}.
\newline\urlprefix\url{https://doi.org/10.1063/5.0046697}

\bibitem{George2021A}
G.~George, L.~Klochko, A.~N. Semenov, J.~Baschnagel, J.~P. Wittmer,
  \href{https://doi.org/10.1140/epje/s10189-020-00004-7}{Ensemble fluctuations
  matter for variances of macroscopic variables}, The European Physical Journal
  E 44~(2) (2021) 13.
\newblock \href {http://dx.doi.org/10.1140/epje/s10189-020-00004-7}
  {\path{doi:10.1140/epje/s10189-020-00004-7}}.
\newline\urlprefix\url{https://doi.org/10.1140/epje/s10189-020-00004-7}

\bibitem{George2021B}
G.~George, L.~Klochko, A.~N. Semenov, J.~Baschnagel, J.~P. Wittmer,
  \href{https://doi.org/10.1140/epje/s10189-021-00070-5}{Fluctuations of
  non-ergodic stochastic processes}, The European Physical Journal E 44~(4)
  (2021) 54.
\newblock \href {http://dx.doi.org/10.1140/epje/s10189-021-00070-5}
  {\path{doi:10.1140/epje/s10189-021-00070-5}}.
\newline\urlprefix\url{https://doi.org/10.1140/epje/s10189-021-00070-5}

\bibitem{Nielsen1996}
J.~K. Nielsen, J.~C. Dyre,
  \href{https://doi.org/10.1103/physrevb.54.15754}{Fluctuation-dissipation
  theorem for frequency-dependent specific heat}, Physical Review B 54~(22)
  (1996) 15754--15761.
\newblock \href {http://dx.doi.org/10.1103/physrevb.54.15754}
  {\path{doi:10.1103/physrevb.54.15754}}.
\newline\urlprefix\url{https://doi.org/10.1103/physrevb.54.15754}

\bibitem{2013i}
\href{https://www.sciencedirect.com/science/article/pii/B9780123870322000131}{Theory
  of simple liquids}, in: J.-P. Hansen, I.~R. McDonald (Eds.), Theory of Simple
  Liquids (Fourth Edition), fourth edition Edition, Academic Press, Oxford,
  2013, p.~i.
\newblock \href
  {http://dx.doi.org/https://doi.org/10.1016/B978-0-12-387032-2.00013-1}
  {\path{doi:https://doi.org/10.1016/B978-0-12-387032-2.00013-1}}.
\newline\urlprefix\url{https://www.sciencedirect.com/science/article/pii/B9780123870322000131}

\bibitem{1980iv}
\href{https://www.sciencedirect.com/science/article/pii/B9780080570464500038}{Copyright},
  in: L.~LANDAU, E.~LIFSHITZ (Eds.), Statistical Physics (Third Edition), third
  edition Edition, Butterworth-Heinemann, Oxford, 1980, p.~iv.
\newblock \href
  {http://dx.doi.org/https://doi.org/10.1016/B978-0-08-057046-4.50003-8}
  {\path{doi:https://doi.org/10.1016/B978-0-08-057046-4.50003-8}}.
\newline\urlprefix\url{https://www.sciencedirect.com/science/article/pii/B9780080570464500038}

\bibitem{Ruscher2017}
C.~Ruscher, A.~N. Semenov, J.~Baschnagel, J.~Farago,
  \href{https://doi.org/10.1063/1.4979720}{Anomalous sound attenuation in
  voronoi liquid}, The Journal of Chemical Physics 146~(14) (2017) 144502.
\newblock \href {http://dx.doi.org/10.1063/1.4979720}
  {\path{doi:10.1063/1.4979720}}.
\newline\urlprefix\url{https://doi.org/10.1063/1.4979720}

\bibitem{CAVAGNA200951}
A.~Cavagna,
  \href{https://www.sciencedirect.com/science/article/pii/S0370157309001112}{Supercooled
  liquids for pedestrians}, Physics Reports 476~(4) (2009) 51--124.
\newblock \href
  {http://dx.doi.org/https://doi.org/10.1016/j.physrep.2009.03.003}
  {\path{doi:https://doi.org/10.1016/j.physrep.2009.03.003}}.
\newline\urlprefix\url{https://www.sciencedirect.com/science/article/pii/S0370157309001112}

\bibitem{Kauzmann1948}
W.~Kauzmann, \href{https://doi.org/10.1021/cr60135a002}{The nature of the
  glassy state and the behavior of liquids at low temperatures.}, Chemical
  Reviews 43~(2) (1948) 219--256.
\newblock \href {http://dx.doi.org/10.1021/cr60135a002}
  {\path{doi:10.1021/cr60135a002}}.
\newline\urlprefix\url{https://doi.org/10.1021/cr60135a002}

\bibitem{Perego2020}
A.~Perego, F.~Khabaz,
  \href{https://doi.org/10.1021/acs.energyfuels.0c01183}{Thermodynamics,
  dynamics, and rheology of fuel surrogates: Application of the
  time--temperature superposition principle in molecular dynamics simulations},
  Energy \& Fuels 34~(9) (2020) 10631--10640.
\newblock \href {http://dx.doi.org/10.1021/acs.energyfuels.0c01183}
  {\path{doi:10.1021/acs.energyfuels.0c01183}}.
\newline\urlprefix\url{https://doi.org/10.1021/acs.energyfuels.0c01183}

\bibitem{Yeh2004}
I.-C. Yeh, G.~Hummer, \href{https://doi.org/10.1021/jp0477147}{System-size
  dependence of diffusion coefficients and viscosities from molecular dynamics
  simulations with periodic boundary conditions}, The Journal of Physical
  Chemistry B 108~(40) (2004) 15873--15879.
\newblock \href {http://dx.doi.org/10.1021/jp0477147}
  {\path{doi:10.1021/jp0477147}}.
\newline\urlprefix\url{https://doi.org/10.1021/jp0477147}

\bibitem{GuevaraCarrion2011}
G.~Guevara-Carrion, J.~Vrabec, H.~Hasse,
  \href{https://doi.org/10.1063/1.3515262}{Prediction of self-diffusion
  coefficient and shear viscosity of water and its binary mixtures with
  methanol and ethanol by molecular simulation}, The Journal of Chemical
  Physics 134~(7) (2011) 074508.
\newblock \href {http://dx.doi.org/10.1063/1.3515262}
  {\path{doi:10.1063/1.3515262}}.
\newline\urlprefix\url{https://doi.org/10.1063/1.3515262}

\end{thebibliography}

\end{document}